\documentclass[epjST,table,usenames,dvipsnames]{svjour}
\usepackage[plain]{fancyref}
\usepackage{graphicx}
\usepackage{tikz}
\usepackage{amsmath,color,amssymb,amsbsy}

\usepackage{url,hyperref}
\hypersetup{colorlinks=true, linkcolor=BrickRed,
  urlcolor=blue!50!black, citecolor=blue!50!black}

\usepackage{pgfplots}

\newcommand{\avg}[1]{\langle #1 \rangle}
\newcommand{\uni}[1]{\hat{\mathbf{#1}}}

\newcommand{\kb}{k_\mathrm{B}}
\renewcommand{\bar}{\overline}
\renewcommand{\vec}[1]{\mathbf{#1}}

\usepackage[sort&compress,numbers]{natbib}

\usepackage{doi}

\definecolor{mypurple}{RGB}{153,61,113}
\definecolor{myblue}{RGB}{63,61,153}
\definecolor{myokker}{RGB}{153,140,61}
\definecolor{mygreen}{RGB}{61,153,86}
\definecolor{mymarine}{RGB}{61,90,153}
\definecolor{mycyan}{RGB}{0,255,255}

\begin{document}
\title{Hot Microswimmers}
\author{Klaus Kroy\inst{1} \fnmsep \thanks{\email{kroy@uni-leipzig.de}}  \and
  Dipanjan Chakraborty
  \inst{2}\fnmsep\thanks{\email{chakraborty@iisermohali.ac.in}} \and
  Frank Cichos \inst{3}\fnmsep\thanks{\email{cichos@uni-leipzig.de}}
}
\institute{Institut f\"ur Theoretische Physik, Fakult\"at f\"ur
  Physik and Geowissenschaften, Universit\"at Leipzig, Leipzig,
  Germany  \and Department of Physical Sciences, Indian Institute of
  Science Education and Research Mohali, Sector 81, SAS Nagar, Manauli
  PO, India \and Institut f\"ur Experimentelle Physik I, Fakult\"at f\"ur
  Physik and Geowissenschaften, Universit\"at Leipzig, Leipzig,
  Germany}
\abstract{ Hot microswimmers are self-propelled Brownian particles
  that exploit local heating for their directed self-thermophoretic
  motion.  We provide a pedagogical overview of the key physical
  mechanisms underlying this promising new technology.  It covers the
  hydrodynamics of swimming, thermophoresis and -osmosis, hot Brownian
  motion, force-free steering, and dedicated experimental and
  simulation tools to analyze hot Brownian swimmers. }
%end of abstract
%
\maketitle

\section{Introduction}
All physicists know Newton's first law of motion, namely that (to
inertial observers) all things move at a constant velocity $\vec v$,
unless acted upon by an external force $\vec F$. It certainly holds
also for particles immersed in a viscous fluid of shear viscosity
$\eta$, at a fundamental level. Yet, if you do not see the solvent,
e.g.\ because it is transparent, the world of these particles looks
very much ``Aristotelian'', in the sense that things usually do not
move without forcing. The reason is that momentum is constantly
dissipated to the solvent (assumed to be at rest).  The law of motion
of the particles therefore depends both on the particle and on the
solvent properties, the best-known example being Stokes' law, $\vec
F=6\pi\eta a\vec v$, for the forced stationary motion of a spherical
bead of radius $a$. Now, turn a small colloidal particle into a
microswimmer, e.g.\ by coating it asymmetrically with a gold cap and
heating it by laser light \cite{jiang-yoshinaga-sano:2010}.  Then it
does move without external forcing. Swimming is a form of autonomous
motion.  Axially symmetric swimmers, the simplest examples for active
colloids, do seem to obey Newton's first law. They are in uniform
force-free motion, despite being immersed in a dissipative solvent.
However, in contrast to Newton's particles in vacuum, their uniform
drift arises from a complex dissipative solvent flow around the
particle and relies on a constant supply of energy, e.g.\ in the form
of chemical fuel, acoustic actuation, or laser heating, that maintains
the system far from equilibrium \cite{cates:2012}. Moreover, the
solvent motion never is entirely deterministic but introduces Brownian
fluctuations in the position and orientation of the particle that will
sooner or later randomize the particle path.  In the following, we
gather some facts and ideas about how such Brownian swimming motion
actually comes about and how it is quantitatively described.  Within
the broad class of artificial micro- and nano-swimmers, which can
nowadays be fabricated in large numbers and with a great variety of
propulsion mechanisms \cite{sengupta-ibele-sen:2012}, we focus on
so-called hot Brownian swimmers --- thermally anisotropic Brownian
particles fueled by (optical) heating.  Briefly, the propulsion of
heated particles works as follows. First, a geometric/material
asymmetry of the particle is exploited to establish an asymmetric
temperature profile in the surrounding solvent upon particle heating.
There ensues a thermoosmotic flow along the surface of the particles.
It can either be harnessed for pumping, if the particle is held fixed
in space by some external force, or for phoretic self-propulsion, if
the particle is mobile.  Boiling is strongly suppressed by the Laplace
pressure (which is inversely proportional to the radius of curvature
and can therefore be quite substantial for small particles) and
heating is highly localized for nanoparticles. Therefore,
substantially larger temperature gradients and, accordingly, more
efficient thermophoretic transport can be realized with
self-thermophoresis than with conventional macroscopic thermophoresis.
Self-thermophoresis\index{self-thermophoresis} is thus arguably an
interesting and technologically promising propulsion mechanism, with
some important advantages over other designs. \emph{i) Universality,
  Availability, Biocompatibility:} it does not rely on exotic (maybe
poisonous) solvents or fuels, but exploits a comparatively
``universal'' mechanism. It does not run out of fuel and is minimally
invasive, since the heating is local and sizable motion can already be
achieved with minor heating of the surroundings.  \emph{ii) Control:}
the propulsion speed can be regulated continuously and propulsion can
instantly be switched on and off, e.g.\ by using conventional lasers
and microscopy equipment. Thanks to emerging efficient cooling
mechanisms for colloidal particles \cite{roder-etal:2015}, one can
even imagine fabricating particles with a reverse gear.  Besides,
efficient force-free steering mechanisms such as photon nudging
\cite{bregulla-yang-cichos:2014} are already available.  \emph{iii)
  Versatility:} heating can be realized by a variety of methods, such
as the absorption of laser light by metal or carbon parts, or of
microwaves by super-paramagnetic parts, which opens up the possibility
of combining several independent and independently addressable
thermophoretic propulsion mechanisms into one microstructure.  A whole
community of researchers is moreover interested in micro- and
nano-particle heating for its own sake \cite{govorov-richardson:2007}.
\emph{iv) Scalability:} downscaling does not reduce the propulsion
speed \cite{bregulla-yang-cichos:2014} but increases the efficiency
\cite{cichos2015fd2015}. Synthesis, speed control, and steering
\cite{bregulla-yang-cichos:2014} of self-thermophoretic swimmers, as
well as their photothermal detection \cite{selmke2013twin}, are all
scalable to nanoscopic dimensions.

The remainder of the paper is structured as follows. The next section
reviews the distinction between self-phoretic, phoretic, and passive
motion and the concomitant flow fields excited in the solvent,
\emph{on a hydrodynamic level}. This means that the physical mechanism
that actually drives the motion is confined to such a narrow boundary
layer that it is sufficiently well captured by a mere hydrodynamic
boundary condition for the solvent flow. On this level, the theory of
swimming is universal and independent of the actual propulsion
mechanism. The following section provides a closer look at the engine
of the swimmer, namely the processes of phoresis and self-phoresis of
heated particles. The basic principles of self-thermophoretic
propulsion and the underlying osmotic processes are very similar to
those in other phoretic phenomena, such as diffusiophoresis and
electrophoresis.  In fact, in real-world realizations, one often
encounters a complex mix of several such mechanisms, which can be hard
to disentangle for small particles in water, which is arguably the
most interesting system for many applications. The particles are
usually charged and surrounded by counter-ions and dissolved salts,
all effects being sensitive to temperature.  And, more often than not,
everything is observed close to a surface, to which similar
considerations apply. Next, we also consider the fluctuating part of
the motion of a hot Brownian swimmer, namely its hot Brownian
motion. This complex nonequilibrium motion limits the deterministic
swimming motion by randomizing the swimming direction and therefore
needs to be understood.  With regard to the mentioned experimental and
conceptual difficulties encountered when working with hot Brownian
simmers, the ability to numerically simulate models of reduced
complexity is particularly valuable. We therefore provide a brief
overview over a nonequilibrium molecular dynamics approach that is
suitable to simulate hot nano-swimmers on a coarse-grained, yet
atomistic basis.  Finally, we discuss some dedicated experimental
techniques for the detection (photothermal imaging and correlation
spectroscopy) and force-free steering (photon-nudging) of hot
swimmers.

\section{Hydrodynamics: Dragging, Swimming, Phoresis}
The contents of this section have in principle been known for more
than a century, explicitly at least for some decades, and something
similar holds for  Sec.~\ref{sec:thermoosmosis} (see
e.g.\ \cite{lighthill:1952,anderson:1989}). So almost everything has
been said, though not yet by everybody. This lack of eloquence has
been cured by the more recent literature, which moreover has stirred
up some confusion by muddling with the distinction between dragging,
swimming, and phoresis (i.e., some writers do not bother to distinguish
sail boats from motor boats). This motivated us to include this pedagogic
material here, although there are already a number of recent technical
reviews clarifying these classical topics for interested contemporary
readers (e.g.\ \cite{juelicher-prost:2009,wuerger:2010,poon:2013}).

For any small suspended particles moving at slow speeds --- technically
speaking, at low Reynolds numbers $|\vec v| a/\nu\ll 1$ --- the
spatio-temporal solvent velocity field $\vec u(\vec r,t)$ reacts
basically instantaneously in the region where it takes any sizable
values, and therefore follows
\cite{veysey-goldenfeld:2007} from the stationary Stokes equations\index{Stokes
  equations},
\begin{equation}
  \label{eq:stokeseq}
   \nu \nabla^2 \vec u = \nabla P  - \vec f\;,\qquad
   \nabla\cdot \vec u =0\;,
\end{equation}
Here, $\nu\equiv \eta/\rho$ denotes the kinematic viscosity, $\rho$ is
the mass density of the solvent, and $\vec f$ an optional external
force density. The hydrodynamic ``pressure'' $P$ (actually $P$ is the
pressure divided $\rho$) can be thought of as a Lagrange multiplier
that serves to account for the second equation. The latter asserts
that (without imposed hydrostatic pressure gradients) stationary flows
have constant density and therefore must be divergence-free to respect
mass conservation: $0\equiv \partial_t\rho=-\rho\nabla\cdot \vec u$.
Equation~(\ref{eq:stokeseq}) is the universal basis for a hydrodynamic
description of passive and active colloidal particles. On a
coarse-grained phenomenological level (i.e.\ not looking too closely
into what is actually really happening very near to the particle
surface), passive particles and swimmers merely differ by the
hydrodynamic boundary conditions imposed on Eq.~(\ref{eq:stokeseq}) at
the particle surface.  The generic boundary condition on solid
surfaces is $\vec u|_{\cal S} =\vec v$, i.e., the velocity of the
fluid and particle match at the particle surface (``no slip, no
influx'').  Yet, allowing the solvent to slip at the particle surface
is the easiest way for a theoretician to turn a passive particle into
a swimmer --- and usually a very good coarse-grained model for
artificial phoretic swimmers and all kinds of animalcules propelled by
their multi-ciliated or dynamically wrinkling skins.  Boundary
conditions and flow fields corresponding to passive drag and active
swimming can simply be superimposed by virtue of the linearity of
Eq.~(\ref{eq:stokeseq}).

Two further remarks seem in order, here. First, note that
the solutions of  Stokes' equations depend on the density and
viscosity of the fluid only in the combination of the kinematic
viscosity $\nu= \eta/\rho$. This has an interesting
consequence. Namely, while air and water feel very differently to
human swimmers because of their large density ratio ($\simeq 10^3$)
and the high Reynolds numbers involved ($\simeq 10^6$), their
kinematic viscosities are on the same order; hence, microswimmers
actually would not mind the difference. Further, note that $\nu$ has
the physical units of a diffusion coefficient and does indeed describe
the diffusion of vorticity through the fluid, which is how momentum
gets dispersed. To see this, and that the stationarity assumption
amounts to the idealization of an infinitely fast diffusion of the
vorticity $\nabla\times \vec u$, simply take the curl of Eq.~(\ref{eq:stokeseq}).

\begin{figure}
\includegraphics[width=0.26\hsize]{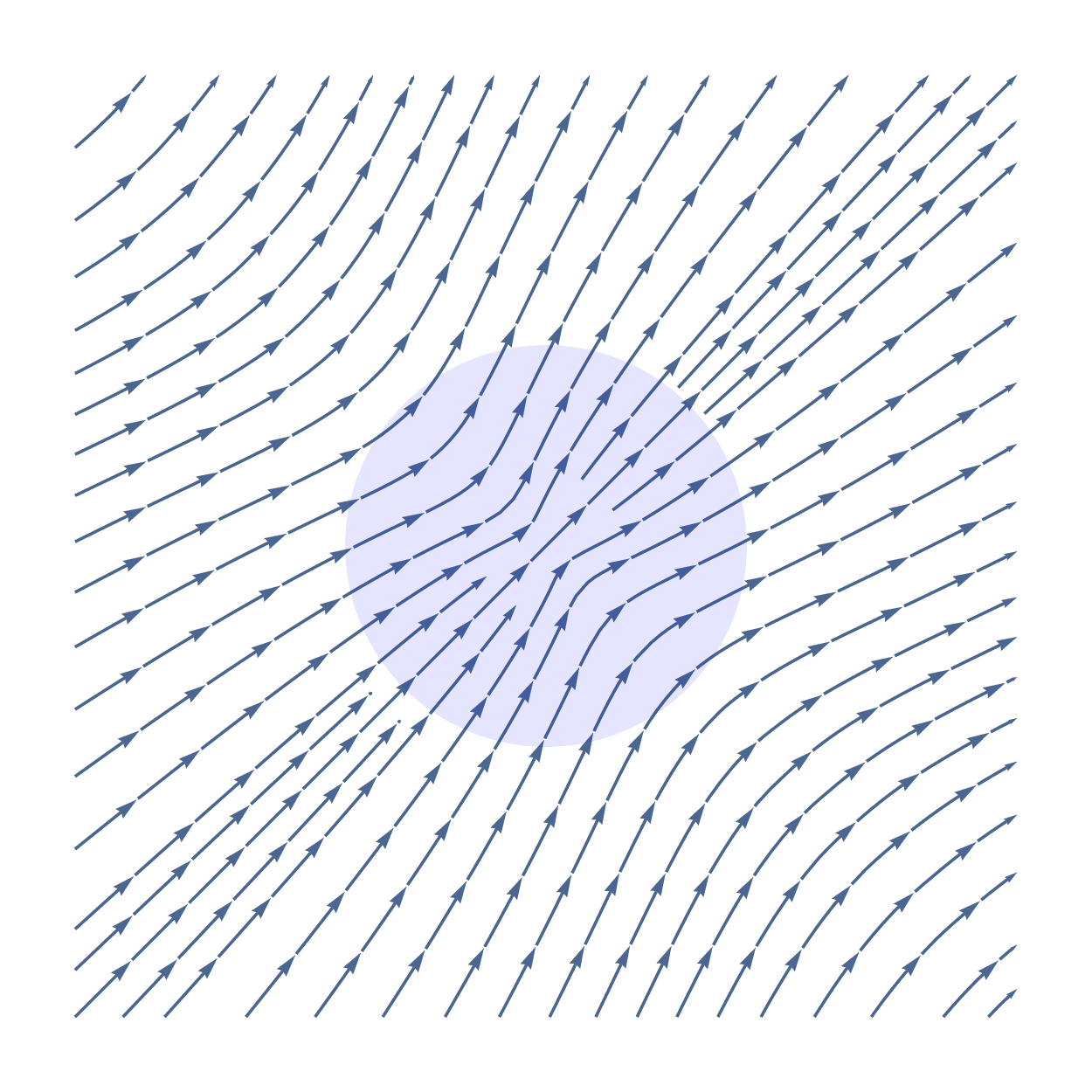}\!\!\!\!\!\!\!\!\!\!\!\!\!(a)
\includegraphics[width=0.26\hsize]{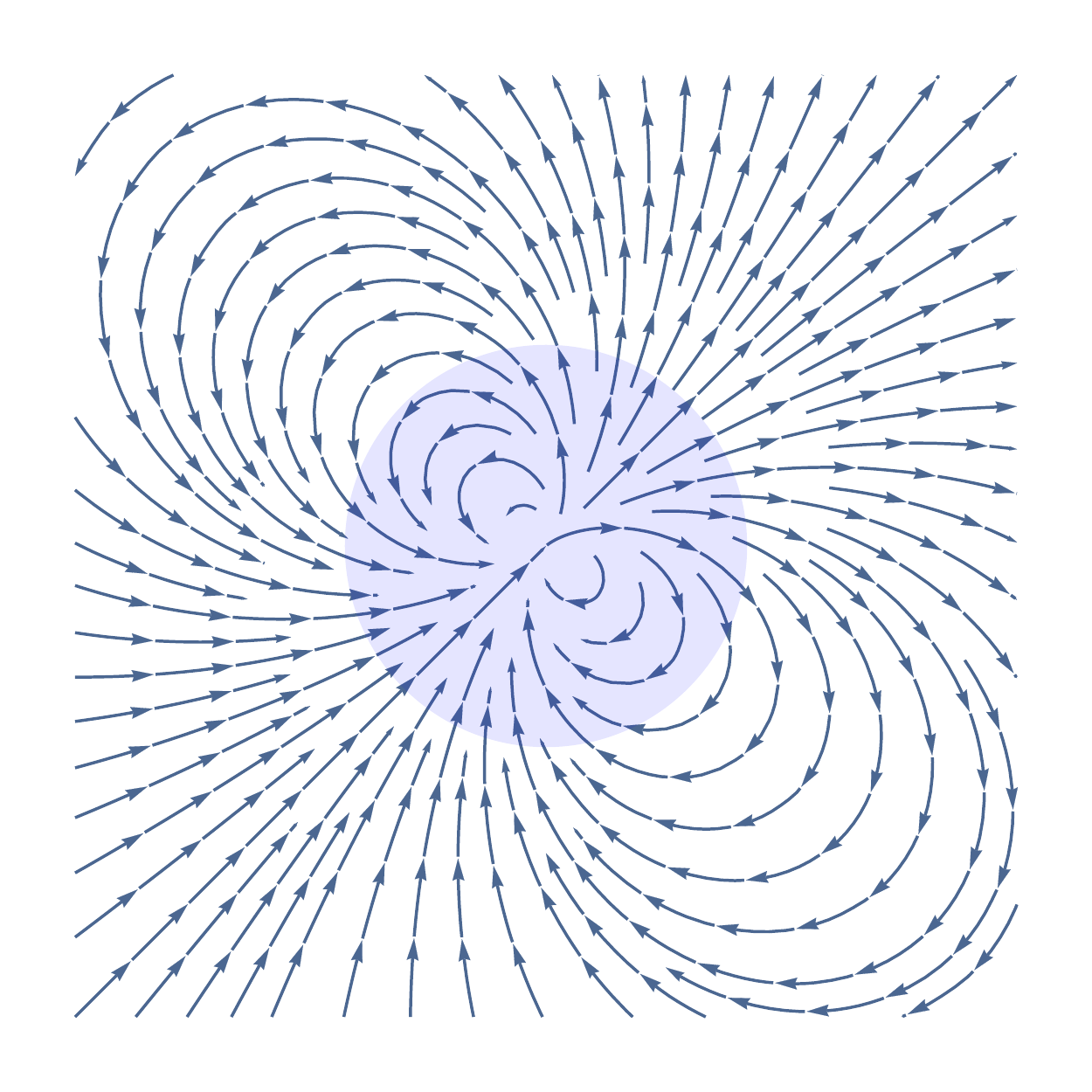}\!\!\!\!\!\!\!\!\!\!\!\!\!(b)
\includegraphics[width=0.26\hsize]{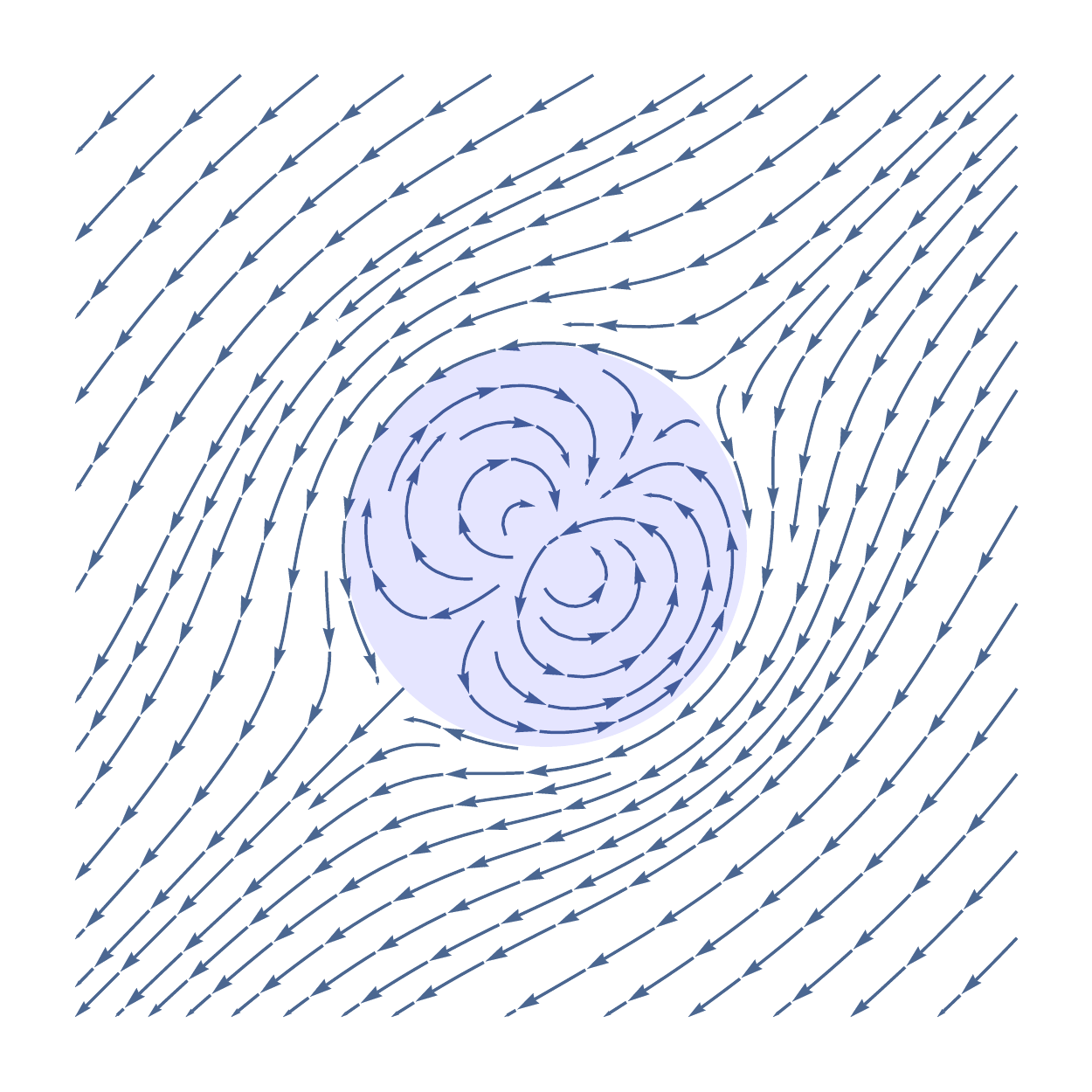}\!\!\!\!\!\!\!\!\!\!\!\!\!(c)
\includegraphics[width=0.26\hsize]{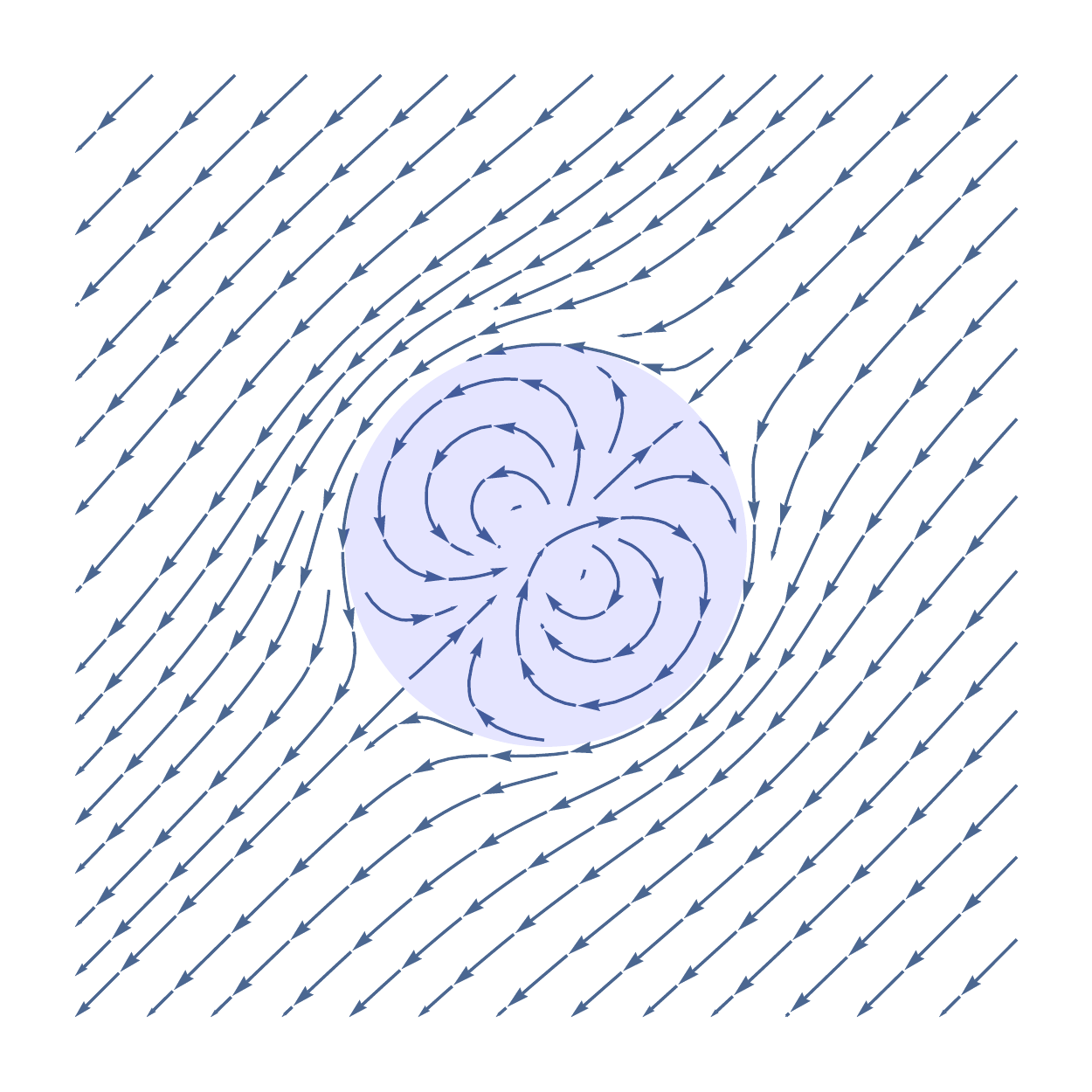}\!\!\!\!\!\!\!\!\!\!\!\!\!(d)
\caption{Streamlines for the contributions to the Stokes flow around a
  sphere, given in Eq.~(\ref{eq:stokesflow}), the ``stokeslet'' with
  $\vec u\propto r^{-1}$ (a) and the ``sliplet'' with $\vec u\propto
  r^{-3}$ (b). The representation in a co-moving frame shows that the
  combination of stokeslet and sliplet in Eq.~(\ref{eq:stokesflow})
  ``sticks'' to, and forcefully pushes through the fluid (c), while
  the sliplet alone slyly ``sneaks'' through the fluid (d). Phoretic
  motion along a perfectly linear thermodynamic gradient and perfect
  (``neutral'') swimmers are phenomenologically described by a pure
  sliplet. Most actual realizations of swimmers, such as bacteria,
  sperms, or (self-)phoretic particles, are not perfect and excite
  additional (undesired) fluid motion,
  cf.\ Fig.~\ref{fig:streamlines2}.  (Note that the direction of
  motion is aligned with the diagonal in all plots.)}
  \label{fig:streamlines}
\end{figure}

\textbf{Passive Transport\index{passive transport} or
  Drag\index{drag}.}  Consider a colloidal sphere of radius $a$ moving
in a liquid solvent of shear viscosity $\eta$. The law relating the
particle's velocity $\vec v$ to the applied external force $\vec F$
(the analog of Newton's second law for a dissolved particle) is
Stokes' law\index{Stokes' law},
\begin{equation}
  \label{eq:stokeslaw}
  \vec v = \frac{\vec F}{ 6\pi\eta a } \;.
\end{equation}
Stokes derived it in the mid-19th century from
Eq.~(\ref{eq:stokeseq}), which yields the flow field
\begin{equation}
  \label{eq:stokesflow}
    \vec u(\vec r) = \frac{3a}{4r}(\vec 1+\uni r \uni r) \cdot \vec v
    + \frac{a^3}{4r^3}(\vec 1-3\uni r \uni r) \cdot \vec v
\end{equation}
around the sphere, where $\uni r \equiv \vec r/r$. This important
result is illustrated in Fig.~\ref{fig:streamlines}.

In the far field (or for an infinitely small sphere), the flow
velocity in Eq.~(\ref{eq:stokesflow}) is dominated by the first term,
with the reciprocal dependence on the distance $ r\equiv |\vec r|
$. Using Eq.~(\ref{eq:stokeslaw}), it is immediately recognized as the
so-called force monopole or ``stokeslet\index{stokeslet}'',
\begin{equation}
  \label{eq:stokeslet}
  \vec u(\vec r) = \frac{1}{8\pi \eta  r}(\vec 1+\uni r \uni r)\cdot \vec F
  \;,\qquad  r \gg a \;,
\end{equation} 
which is the fundamental solution of Eq.~(\ref{eq:stokeseq}), i.e.,
the fluid response to a point force $\vec f = \vec F\delta(\vec r)$
applied at the origin (Fig.~\ref{fig:streamlines}a).  It is solely
responsible for dispersing all the momentum and vorticity supplied to
the fluid by the drag force $\vec F$, as required by Newton's third
law, and apparent from its reciprocal dependence on the distance. (A
fluid velocity $\vec u\propto r^{-1}$ amounts to a total momentum
leaking out to infinity at a constant rate --- i.e., perfect force
balance with the surroundings.)  Accordingly, Eq.~(\ref{eq:stokeslet})
alone reproduces Stokes' law, Eq.~(\ref{eq:stokeslaw}), upon averaging
over the surface of the sphere at $r=a$. Indeed, the projector in
parentheses in Eq.~(\ref{eq:stokeslet}) renders $\vec u$
divergence-free and averages to $\avg{\vec 1+\uni r \uni r}=\vec
1\text{tr} (\vec 1+\uni r \uni r)/3=\vec 1(3+1)/3=(4/3)\vec 1$, as
required. In summary, stokeslet and external force are inseparable
twins that always come and go together, you cannot have one without
the other.

\begin{figure}
\includegraphics[width=0.26\hsize]{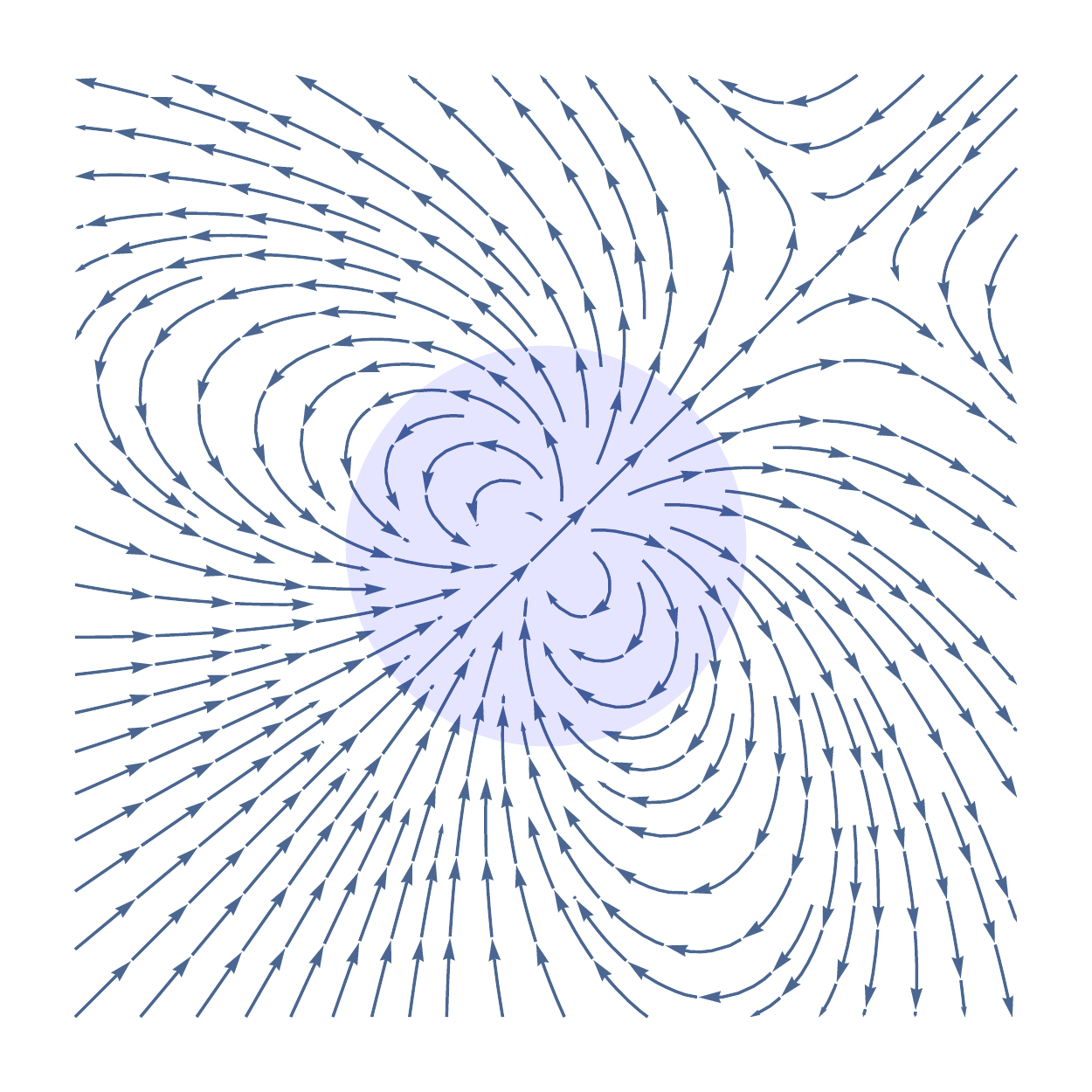}  \!\!\!\!\!\!\!\!\!\!\!\!\!(a)
\includegraphics[width=0.26\hsize]{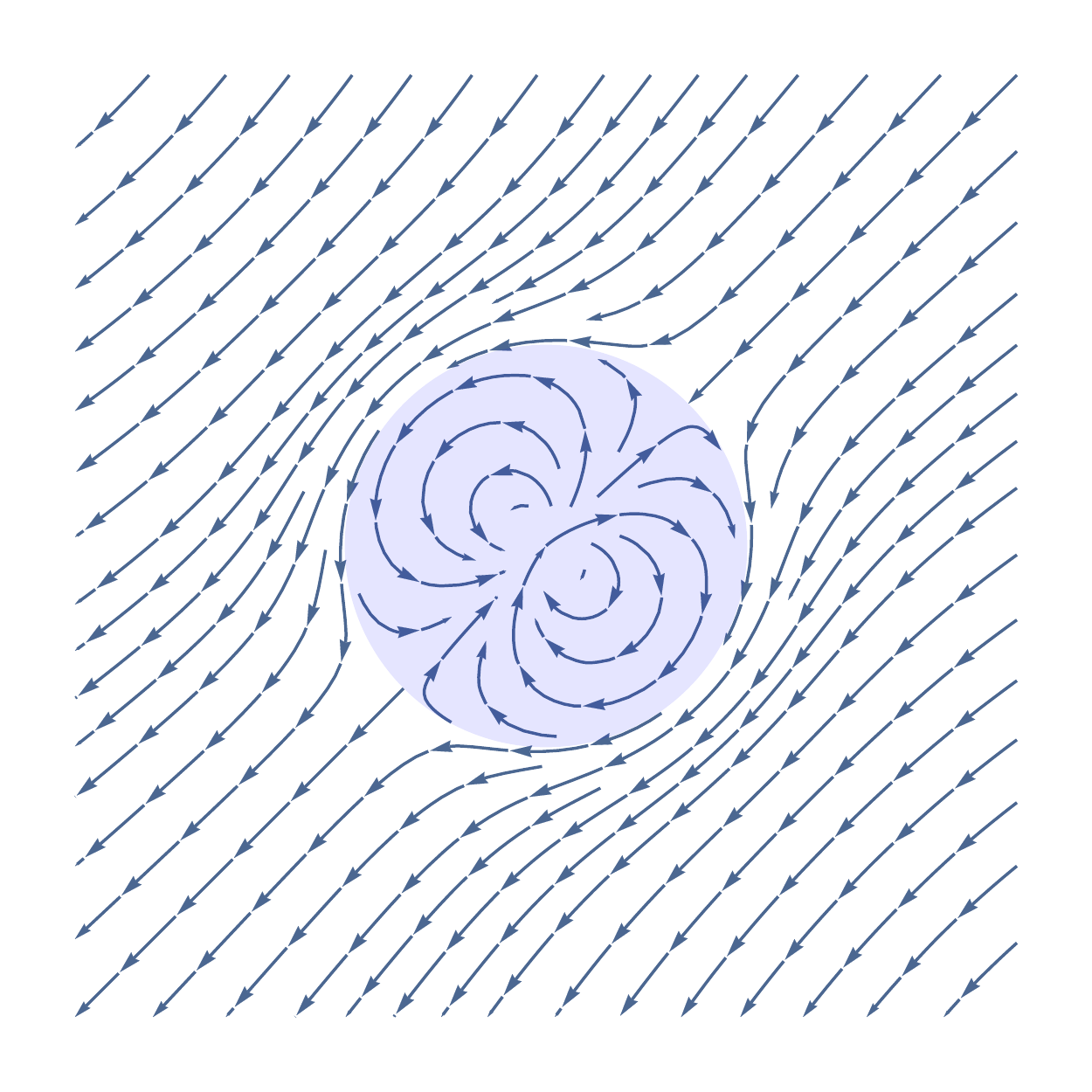} \!\!\!\!\!\!\!\!\!\!\!\!\!(b)
\includegraphics[width=0.26\textwidth]{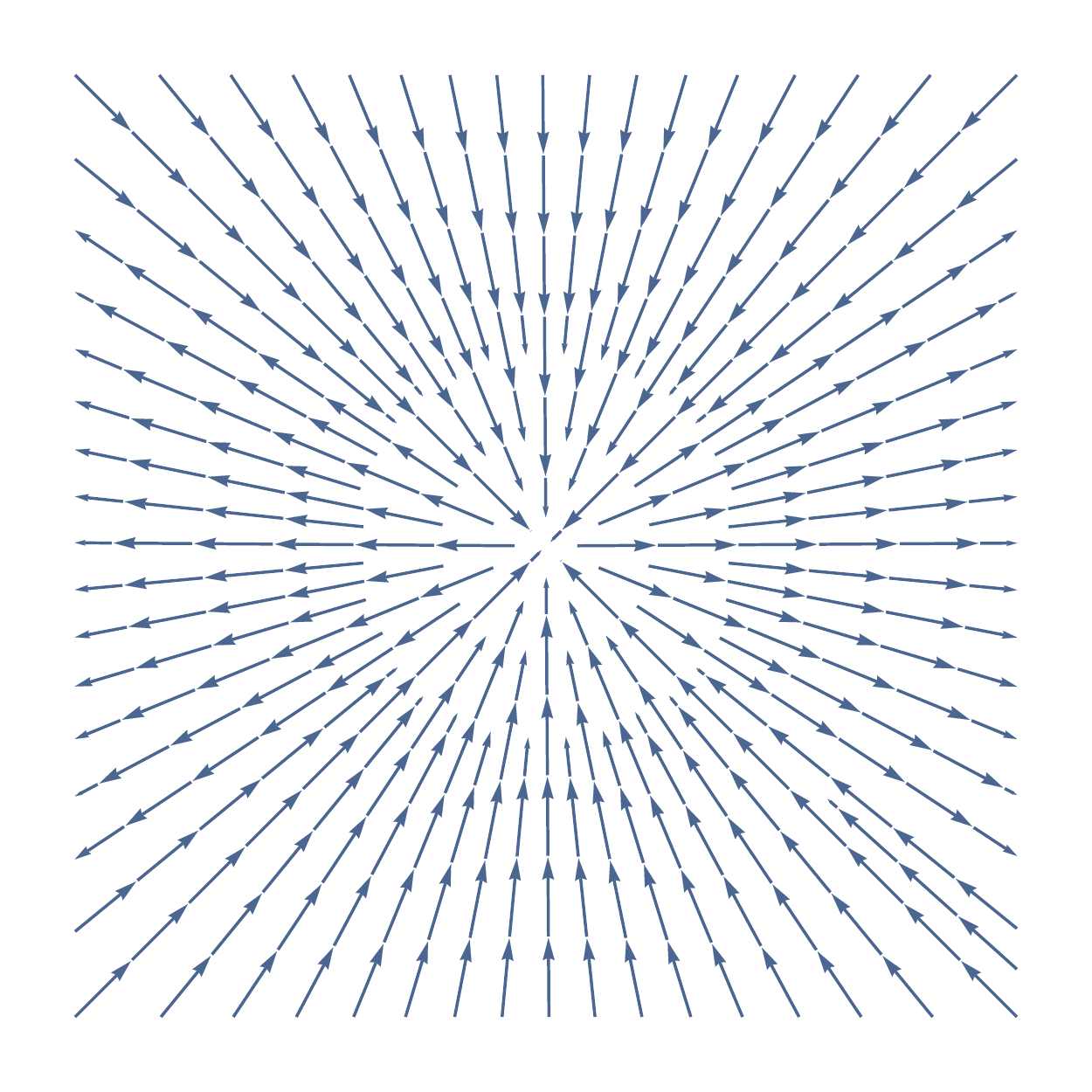}\!\!\!\!\!\!\!\!\!\!\!(c)
\includegraphics[width=0.24\textwidth]{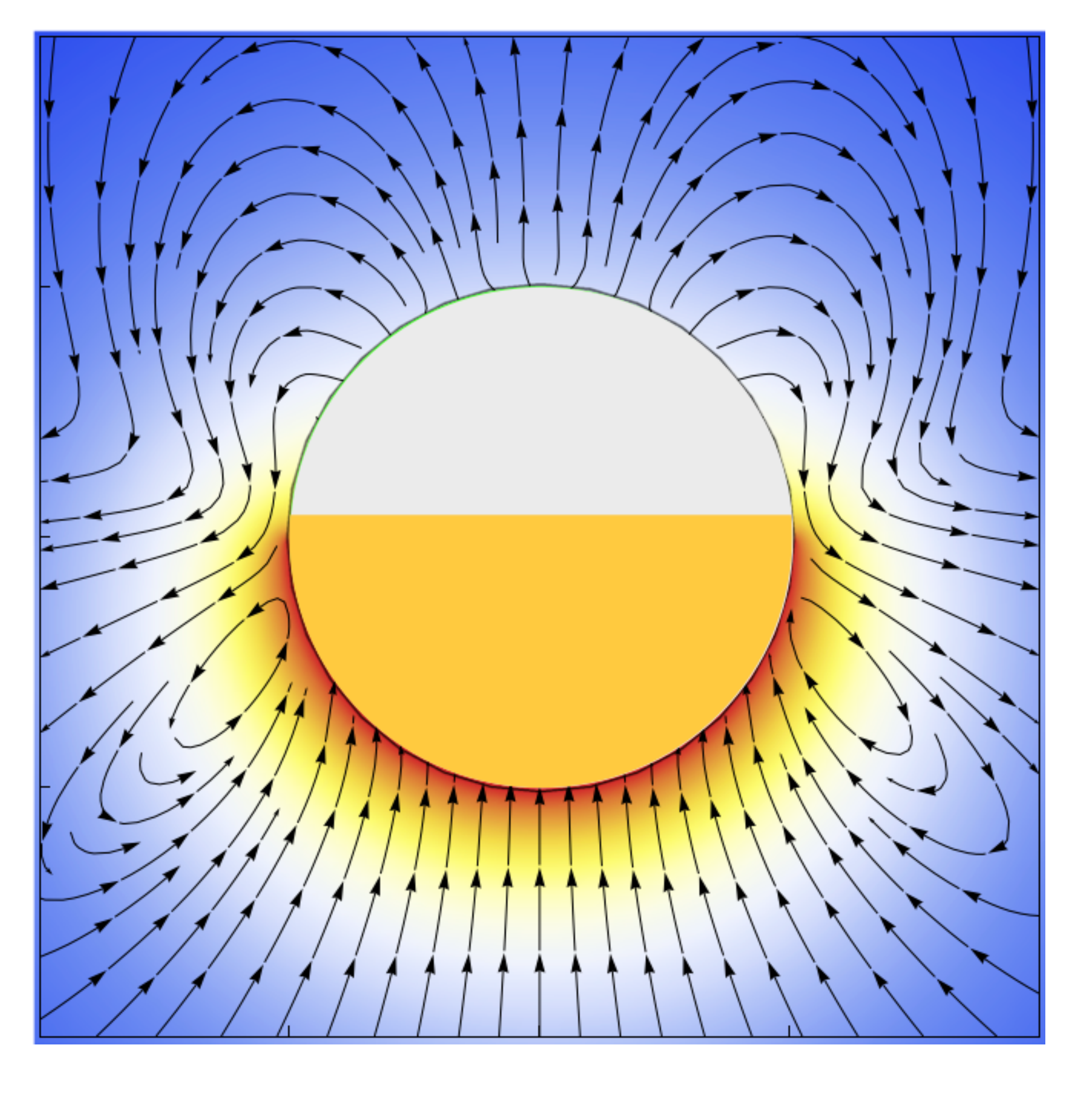}\!\!\!\!\!\!\!\!\!\!(d)
\caption{Lab-frame streamlines for an imperfect squirmer,
  consisting of a ``sliplet'' $\vec u\propto r^{-3}$ (the
  perfect engine) and a stokeslet dipole $\vec u\propto r^{-2}$ (the
  dominant far-field signature of imperfections) in force-free swimming mode
  (a) and in pumping mode (b), which creates a stokeslet that
  counterbalances the external stalling force keeping the pump in
  place. The forth-back/left-right symmetric dipole field alone (c)
  corresponds to the far field of a puller/pusher moving
  vertically/horizontally, respectively, whereas the analytically
  calculated streamlines around a more realistically modeled
  force-free Janus particle with a heated gold cap (d, adapted
  from \cite{bickel-majee-wuerger:2013}) exhibits pronounced
  near-field structure.}
  \label{fig:streamlines2}
\end{figure}

The stokeslet does not exactly conform with the no-slip boundary
condition, though. In fact, it does not even bear any signature of the
particle size, at all. This is where the second term in
Eq.~(\ref{eq:stokesflow}) comes in, which is known as a source
doublet. Its sole purpose is to let the velocity slip along the
particle surface, so that it could also very well be called a
``sliplet''. By construction, it precisely corrects the ``failure'' of
the stokeslet, Eq.~(\ref{eq:stokeslet}), to account for the finite
size of the particle. This statement is easily verified by setting
$r=a$ in Eq.~(\ref{eq:stokesflow}) and gleaned from
Fig.~\ref{fig:streamlines}, which depicts the streamlines around a
dragged particle according to Eq.~(\ref{eq:stokesflow}) (c), and the
sliplet alone (d), in the particle frame. The figure moreover shows
that the dragged particle perturbs the surrounding solvent much more
than a particle motion that involves only a sliplet.  Note, as an
aside, that Figs.~\ref{fig:streamlines}, \ref{fig:streamlines2} also
indicate that the visual appearance of the flow field changes
drastically with the reference frame, while subtleties related to the
character of the flow field may be hard to discern, which complicates
experimental studies of the hydrodynamics of low-Reynolds-number
swimming \cite{drescher-etal:2011}.

\textbf{Phoresis\index{phoresis}.} According to the above discussion,
the second term in Eq.~(\ref{eq:stokesflow}), the source doublet or
sliplet, does not change the average fluid velocity on the particle
surface, nor does it carry vorticity that would give rise to a net
force or leak momentum to infinity. It idealizes the complex dynamic
processes, which actually take place in a narrow boundary layer around
the particle, as a kind of tank-treading motion.  With its help, the
sphere effectively sneaks through the fluid
(Fig.~\ref{fig:streamlines}d).  The normal velocity component of the
sliplet on the particle surface is $\uni r \cdot \vec u(a)=(1-3)\uni
r\cdot \vec v/4=-(v/2)\cos\theta$ in the lab frame. (For the moving
frame, subtract $\vec v$ with normal and tangential components
$v\cos\theta$ and $v\sin\theta$, respectively.)  This means that the
sliplet alone, without the stokeslet, fulfills the no-influx boundary
condition for a sphere moving with velocity $-\vec v/2$. Therefore,
taking twice the negative sliplet in Eq.~(\ref{eq:stokesflow}), namely
\begin{equation}
  \label{eq:sliplet}
   \vec u = \frac{a^3}{2r^3}(3\uni r \uni r-\vec 1)  \cdot \vec v
\end{equation}
allows a spherical particle to move at velocity $\vec v$ without
violating the no-influx boundary condition. In other words, we have
just constructed a perfect swimmer. For such a perfect swimmer (or
should we say ``slipper''?), the tangential component of the fluid
velocity at the surface contributes $\uni v\cdot (\vec 1 -\uni r \uni
r)\cdot \vec u(a)=-(v/2)\sin^2\!\theta$ along the propulsion direction,
i.e., it must be $-(v/2)\sin\theta$. The relative tangential velocity of
the fluid in the particle frame therefore does not vanish but has a
slip of magnitude $-(3v/2)\sin\theta$, corresponding to a maximum
tangential slip velocity of magnitude $-3v/2$ at the equator. If its
projection $-(3v/2)\sin^2\!\theta$ onto the propulsion direction is
averaged over the sphere, one exactly recovers the negative propulsion
velocity.

Now, observe that a sinusoidal variation of the tangential slip
velocity along the circumference of the particle is precisely what one
would expect if the slip mechanism was caused by the tangential
component of some linear external gradient along the propulsion
direction.  The sliplet is thus not only a theoretician's dream of a
swimmer; it is at the same time the simplest coarse-grained
hydrodynamic description of a homogeneous sphere moving by any kind of
phoretic mechanism, whenever the microscopic processes responsible for
the phoretic slip are confined to a narrow boundary layer at the
sphere's surface and proportional to an undisturbed linear external
gradient.  As an obvious consequence, the swimming velocity of a
force-free phoretic particle must be independent of its size if all other
parameters are held constant. (The actual size dependence of
identically designed hot Janus swimmers exposed to the same laser
power is a more subtle issue, since it depends on the size dependence
of the scattering cross section \cite{bregulla-yang-cichos:2014}).

\textbf{Self-Phoresis\index{self-phoresis}} Self-phoresis is just like
phoresis, the only difference being that the gradient of the relevant
field (the electrical potential in electrophoresis, the concentration
of a solute in diffusiophoresis, the temperature in thermophoresis,
etc.) is created by the particle itself rather than being externally
imposed. Depending on the design of the self-phoretic particle, the
self-generated gradient field varies more or less nonlinearly along
the particle axis, so that hydrodynamic perturbations to the pure
sliplet are usually induced. The leading term, dominating the far
field, is a force (or stokeslet) dipole.  It can be imagined as
consisting of two stokeslets ($\vec u\propto r^{-1}$) of equal
strength that are slightly symmetrically displaced from the center
along the particle axis and pointing in opposite directions along this
axis. The force dipole thus excites a forth-back and left-right
symmetric flow field that decays like $\vec u\propto 1/r^2$
(Fig.~\ref{fig:streamlines2}). It does not contribute to the
propulsion, but, as the slowest radially decaying component in the
flow field, it dominates the interactions of most imperfect swimmers
with walls and other particles or swimmers. Neglecting all
higher-order (faster decaying) corrections amounts to the simplest
``squirmer'' approximation, in which self-phoretic swimmers can be
classified according to the strength and orientation (in or out) of
their stokeslet-dipole term. They are called ``pushers'' or
``pullers'' if the flow created by the stokeslet dipole is oriented
outwards or inwards along the particle axis, respectively, and
``neutral squirmers'', if the stokeslet dipole is absent. Pullers are
naturally attracted to each other and to walls head-on, while pushers
(which actually pull sidewards) tend to align with walls and other
particles or swimmers \cite{llopis-pagonabarraga:2010}. In any case,
because of the linearity of the flow, the hydrodynamic effect of walls
is very much like the effect of other swimmers. In simple geometries
it can be simulated by image swimmers, as familiar from the
image-charge method in electrostatics.  The pertinence of the simple
squirmer description with only a sliplet and a dipole term
(Fig.~\ref{fig:streamlines2}a) can be gauged by comparison with the
plot in Fig.~\ref{fig:streamlines2}d, which depicts the exact solution
for the thermoosmotic flow around a Janus swimmer with a hot
isothermal gold cap, assuming (identical) finite thermal
conductivities of bulk and solvent and, of course, an infinitely thin
thermoosmotically active boundary layer
\cite{bickel-majee-wuerger:2013}. While some overall resemblance can
be detected if the different propulsion directions (diagonal/upwards)
in the two plots are taken into account, the more realistic flow field
is obviously strongly affected by additional near-field
contributions. Recent attempts to classify the collective behavior in
dense suspensions by simplified hydrodynamic models might therefore
have to be taken with a grain of salt.

\section{Propulsion: Thermoosmosis\index{Thermoosmosis} and
  Thermophoresis\index{Thermophoresis}}\label{sec:thermoosmosis}

\begin{figure}
    \centering
     \includegraphics[width=0.8\hsize]{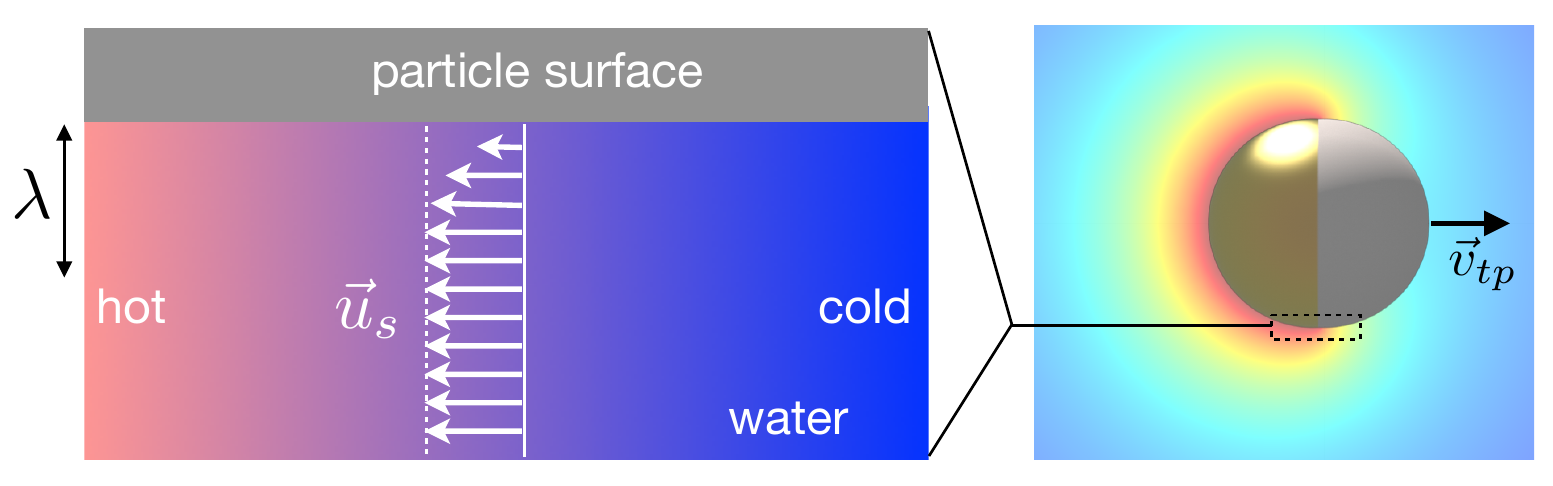}
     \caption{An externally applied temperature gradient along a
       liquid-solid interface excites a thermoosmotic creep
       flow. Hydrodynamically, i.e., if seen from a distance much
       larger than the interaction length $\lambda$ determined by the
       physics and chemistry at the solid-liquid interface, it can be
       described by a local surface slip velocity $\vec
       u_\text{s}(\vec r_\|)$. Lateral variations can arise from
       lateral variations in the solid-liquid interactions or in the
       temperature gradient.  At a fixed solid boundary, the
       solvent slip acts as an osmotic pump, while it turns a
       dissolved mobile particle into a phoretic swimmer. Its
       thermophoretic drift velocity $\vec v=\vec v_\text{tp}$ is
       given by the surface-average of the negative slip velocity
       $-\vec u_\text{s}(\vec r_\|)$.}\label{fig:figure1}
\end{figure}

As stated in the previous section, a particle with a tangential slip
component of the fluid velocity at its surface would allow to
construct an ideal swimmer.  To achieve this tangential surface
slip one may tune the interfacial interactions between the liquid
and the particle. In the colloidal domain, the interaction range at
the liquid-solid interface usually decays within a certain interaction
distance $\lambda$ from the interface (see Fig.~\ref{fig:figure1}).
Within this boundary layer, the interfacial excess enthalpy $h(r_{\!\perp})$
decays to zero, where $r_{\!\perp}$ is the distance from the interface
\cite{Derjaguin:DokladyAkadNaukSssr:1941}. Depending on the surface
chemistry and on the complexity of the solvent, the interfacial excess
enthalpy may comprise a number of different contributions. Simple
microscopic model descriptions are available for some special cases
such as interfacial electrostatic interactions arising form a charged
solid surface and a counter ion cloud bound in a double layer
\cite{anderson:1989}.

If a thermodynamic field gradient, e.g.\ of an electric field
\cite{anderson:1989}, or of a solute concentration
\cite{Popescu_EPJST}, or temperature
\cite{Derjaguin:DokladyAkadNaukSssr:1941}, with a tangential component
is externally applied, this modifies the thermodynamic balance in the
interfacial layer and causes a tangential interfacial creep flow. In
particular, for a temperature gradient $\vec{\nabla}_{||} T$ along a
container wall, this creep flow is a so-called thermoosmotic flow. The
wall acts as an osmotic pump.  The flow velocity vanishes at the solid
boundary and saturates at a distance around the interaction length
$\lambda$ (see Fig.~\ref{fig:figure1}).  Hydrodynamically, i.e.,
seen from a distance much larger than $\lambda$, the effect can be
described by a local slip velocity $\vec u_{\rm s}(\vec r_\|)\equiv
\mu (\vec{r_\|}) \nabla_{||} T$ of the solvent at the surface. The
thermophoretic mobility $\mu(\vec{r_\|})$ has been expressed as an
integral over the sheared boundary layer
\cite{Derjaguin:DokladyAkadNaukSssr:1941,poon:2013}
\begin{equation}\label{eq:slip_velocity}
\mu (\vec{r_\|}) =  -\frac{1}{\eta
  T} \int_{0}^{\infty}\text{d}r_{\!\perp}\, r_{\!\perp}\, h(r_{\!\perp}, \vec{r_\|}) \,\;.
\end{equation}
It depends on the interfacial properties and will generally change
laterally if the liquid-solid interactions vary along the liquid solid
interface, as typically the case for artificial microswimmers. For
example, in the case of a Janus particle, the value of $\mu$ will
generally differ between the two hemispheres. A thermoosmotic slip flow
along the surface of a mobile particle will set the particle into
thermophoretic motion
\cite{Duhr:PhysRevLett:2006,Parola:EurPhysJE:2004}.  As discussed in
the context of Eq.~(\ref{eq:sliplet}), the acquired thermophoretic
drift velocity $\vec{v}_\text{tp}$ follows by averaging the negative
surface creep flow $-\vec u_\text{s}$ over the particles surface $\cal
S$, namely,
\begin{equation}\label{eq:propulsion_v}
\vec{v}_\text{tp}=-\frac{1}{\cal S}\int\vec{u}_\text{s}(\vec r_\|)\,\text{d}{\cal
  S}=-\frac{1}{\cal S}\int\mu (\vec{r}_\|) \nabla_{||} T \,\text{d}\cal S
\end{equation}
If the temperature gradient is generated by the phoretic particle
itself, the ensuing motion can justly be called
``self-thermophoretic''. However, whenever the particle swims close to
a container wall or other boundaries, it will generally induce some
thermoosmosis there and, in return, pick up additional flow
contributions, so that the overall particle velocity $\vec v$
will differ from the nominal thermophoretic drift velocity $\vec
v_\text{tp}$. The same holds for mutual encounters of swimmers.  The
effect can loosely be thought of as a ``catalysis'' mechanism for the
swimmer's propulsion engine.

\section{Fluctuations: Hot Brownian Motion\index{Hot Brownian motion}}
The study of microswimmers has a long history dating back to the 17th
century, when they were first observed under the microscope, most
notably by the dutch draper Antoni van Leeuwenhoek.  Only much later,
starting with systematic investigations by scientists like Robert
Brown and Adolphe Brogniard in the early 19th century, researchers
slowly became aware of the interference of Brownian motion with
micro-scale swimming, and much of the pioneering work was devoted to
disentangling both effects. So the study of animalcules predated that
of molecules, and what started as an investigation of the former
eventually furnished proof of the existence of the latter
\cite{frey-kroy:2005}. Today we are retracing this path backwards,
from bottom-up.  Brownian motion, which is due to thermal fluctuations
of the molecules of the swimmers' medium and cannot be switched off,
is well understood for isothermal solvents. One might think that it
gets easily outpaced by the directed ballistic motion of swimmers, but
it reappears through the back door of rotational Brownian motion that
randomizes the swimming direction. Moreover, since we are particularly
interested in non-isothermal swimmers, driven by thermophoresis and
thermoosmosis, we have to consider non-isothermal or ``hot'' Brownian
motion \cite{rings-etal:2010}, which is an interesting subject by
itself, and has to be understood if one wants perfect control over hot
swimmers.

The heat emanating from a hot micro- or nano-swimmer has two main
effects. It reduces the friction and increases the thermal
fluctuations around the swimmer and thereby enhances the translational
and rotational Brownian motion of the swimmer.  It turns out that a
major simplification occurs for a coarse-grained description that
holds on long times, where the stationary Stokes approximation in
Eq.~(\ref{eq:stokeseq}) holds for the deterministic solvent flow.
This is called the Markovian limit, since it neglects memory effects
due to the slow dynamics of vorticity diffusion, which are already
present in an accurate description of equilibrium Brownian motion
\cite{li-etal:2010,franosch-etal:2011}, and which considerably
complicate the theory of hot Brownian motion
\cite{falasco-etal:2014}. In the Markovian description of hot Brownian
motion, the non-equilibrium effects can be subsumed into a small
number of effective transport coefficients that can analytically and
explicitly be calculated for sufficiently symmetric swimmer designs:
chiefly, an effective reduced friction coefficient $\zeta_\text{HBM}$ and
an effective Brownian temperature $T_\text{HBM}$. The two quantities
determine the effective diffusivity $D_\text{HBM}$ via a generalized
Sutherland--Einstein relation \cite{Chakraborty:2011},
\begin{equation}\label{eq:gser}
  D_\text{HBM}=\frac{k_BT_\text{HBM}}{\zeta_\text{HBM}} \,.
\end{equation}
To estimate the effective friction coefficient $\zeta_\text{HBM}$, the
equation of state $\eta(T)$ of the solvent needs to be known. It can
often (e.g.\ for water) accurately be represented by a Vogel Fulcher
law, $\eta (T) = \eta_{\infty} \exp[A/(T-T_{VF})]$, from which
explicit predictions for the effective translational and rotational
friction coefficients of a hot sphere can be calculated (see
e.g.\ Refs.~\cite{rings-chakraborty-kroy:2012,oppenheimer-navardi-stone:2016}
and the supporting online materials in Ref.~\cite{Chakraborty:2011}).

Practically and conceptually it is more interesting to understand the
effective temperature that characterizes the thermal agitation of 
non-isothermal Brownian particles. For pedagogic reasons, it is best to first
consider a homogeneous spherical particle that is constantly
maintained at a temperature above the ambient temperature, e.g.,
because it diffuses in a laser focus and absorbs the laser light much
more efficiently than its surrounding solvent. The heating creates in
the solvent a radial temperature field $T(\vec r)$, co-moving with the
particle (since heat diffuses via molecular collisions and therefore
much faster than a colloidal particle). It can then be shown that the
Brownian motion of the hot particle is described by the usual
overdamped Langevin equations of motion with the following effective
temperature for the noise strength
\cite{Chakraborty:2011,falasco-etal:2014}
\begin{equation}
	T_\text{HBM}=\frac{\int \! \phi(\vec r)T(\vec r)\,\text{d}\vec r }{ \int \! 
        \phi(\vec r)\,\text{d}\vec r } \;.
\end{equation}
This noise temperature is determined from the condition that it
characterizes the Brownian motion of the heated diffusing particle as
if it were an equivalent isothermal particle in a fluid of constant
temperature $T_\text{HBM}$. Here $\phi(\vec r)$ is the
so-called dissipation function which depends on the viscosity and the
solvent velocity gradient and weighs the importance of fluctuations at
the diverse local temperatures $T(\vec r)$ according to their relevance for the
agitation of the Brownian particle. Due to the different flow fields
for translational (t) and rotational (r) motion of the particle, this prescription leads to
different effective temperatures for the translational and rotational
Brownian motion, namely  \cite{Chakraborty:2011,rings-chakraborty-kroy:2012,falasco-etal:2014} 
\begin{equation}\label{eq:HBM}
  T^\text{t}_\text{HBM}\approx T_0\left (1+\frac{5}{12}\Delta T \right )\;, \qquad
  T^\text{r}_\text{HBM}\approx T_0\left (1+\frac{3}{4}\Delta T \right )\;,
\end{equation}
where $\Delta T$ is the difference between the solvent temperature at
the particle surface and the ambient temperature $T_0$.  Higher order
terms in $\Delta T$, which involve the effective viscosity
$\eta_\text{HBM}(\Delta T)$, can be calculated but are usually small
in actual applications.  Note that, in contrast to certain effective
temperatures that were recently hotly debated in other areas of
non-equilibrium statistical mechanics, the effective temperatures of
hot Brownian motion are not merely postulated but can systematically
be calculated from an underlying non-isothermal fluctuating
hydrodynamic theory \cite{falasco-kroy:2016}, namely
Eq.~(\ref{eq:stokeseq}) with $\vec f(\vec r)$ representing a
non-isothermal noise force. The latter can be expressed as the
divergence of a fluctuating Gaussian shear stress field with a
covariance proportional to $T(\vec r)$.

While the effective translational temperature is usually the only
quantity that matters for a homogeneous hot sphere and is most easily
experimentally inferred from the translational Brownian fluctuations,
its effect is easily outpaced by active propulsion at late times (see
Ref.~\cite{falasco-etal:2016} for a detailed analysis of the velocity
fluctuations of a hot Brownian swimmer).  Then, it is actually the
effective rotational temperature $ T^\text{r}_\text{HBM}$ that matters
more, because it limits the persistence of the directed
propulsion. The observation of trajectories of a hot swimmer could
thus be used to experimentally infer its rotational hot Brownian
temperature $T_\text{HBM}^\text{r}$. For a very precise comparison
with theory, one would then have to account for deviations from
Eq.~(\ref{eq:HBM}) due to the heterogeneous temperature field around the
hot swimmer, depicted in Fig.~\ref{fig:streamlines2}d. To a good
approximation, it is sufficient to evaluate Eq.~(\ref{eq:HBM}) at the
average temperature of the Janus particle, though.

\begin{figure}
    \centering
     \includegraphics[width=0.9\hsize]{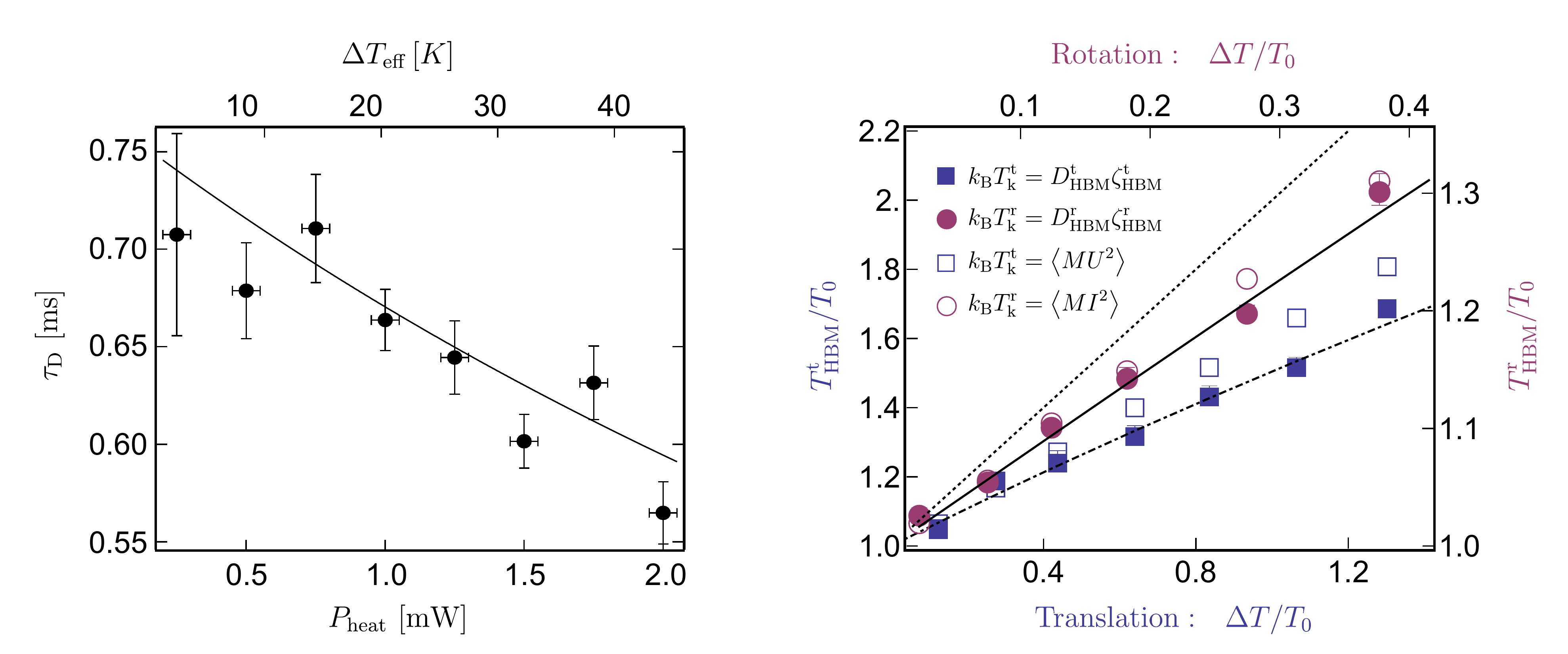}
%\begin{tikzpicture}
%\node at (-50pt, 65pt) {{\it (a)}};
%\node at (120pt, 65pt) {{\it (b)}};
%\node[above,right] (img) at (-2,0)
%{\includegraphics[height=0.38\textwidth]{HBM.pdf}};
%\node[above,right] (img) at (4.,0)
%{\includegraphics[height=0.42\textwidth]{HBM_test2_color.pdf} };
%\end{tikzpicture}
\caption{Parameter-free experimental and numerical tests of the
  predicted hot Brownian diffusivities of a (homogeneous) hot Brownian
  particle, Eq.~(\ref{eq:HBM}). \emph{Left:} In the experiments, the
  Twin-PhoCS method (Sec.~\ref{sec:hotexp}) is employed to measure the
  average time $\tau_D$ for crossing the laser focus
  \cite{selmke2013twin}. \emph{Right:} In the non-equilibrium
  molecular dynamics simulations (Sec.~\ref{sec:hotsim}), the
  effective temperature $T_\text{HBM}$ is deduced via
  Eq.~(\ref{eq:gser}). Lines indicate the solvent temperature at the
  particle surface (dotted) and the theoretical predictions for the
  rotational (solid) and translational (dot-dashed) effective Brownian
  temperatures.  Theoretical predictions \cite{falasco-etal:2014} for
  the effective kinetic temperatures $T_\text{k}^{\text{t,r}}$ (open
  symbols) for the translational (t) and rotational (r)
  \emph{velocities} are not yet available, since the theory has not
  yet been generalized to compressible
  solvents.}\label{fig:figure_HBM}
\end{figure}

So far, several of the predictions of the theory of hot Brownian
motion could be validated experimentally and in numerical simulations
\cite{Chakraborty:2011,rings-etal:2010,selmke2013twin,falasco-etal:2016}.
We specifically mention the experimental verification of the
translational effective temperature from Eq.~(\ref{eq:HBM}).
Interestingly, the experiment can exploit the solvent heating due to
the hot particle for a highly accurate detection of its Brownian
motion, as described in Section~\ref{sec:hotexp}.  Figure
\ref{fig:figure_HBM} (left panel) provides a parameter-free comparison
of the average diffusion time $\tau_{D}$ of a heated Brownian particle
in a laser focus, which has been obtained by this method, with the
prediction in Eq.~(\ref{eq:HBM}). An even more direct comparison of
the various temperatures, i.e., the conventional local molecular
solvent temperature and the effective temperatures characterizing the
Brownian dynamics of various degrees of freedom (rotational,
translational positions and velocities) of the particle is possible in
our atomistic simulations, see Fig.~\ref{fig:figure_HBM} (right
panel).

As a useful and instructive application of these concepts to the
motion a hot Brownian swimmer, we also want to mention the validation
\cite{falasco-etal:2016} of a recently discovered spatial fluctuation
relation \cite{hurtado-etal:2011}. In this work, the fluctuating
velocity of a hot Brownian swimmer was recorded both in experiment and
in numerical simulations. The corresponding histograms were shown to
be in good accord with the predicted fluctuation theorem for the
probabilities $P(\vec J)$ to observe particle currents $\vec J$, $\vec
J'$ of equal strength $J$ but in different directions, namely,
\begin{equation}\label{eq:FR}
P(\vec J)=P(\vec J') e^{{\cal F}  \cdot (\vec J - \vec J')} \,.
\end{equation}
Here, the strength $|{\cal F}| = v \zeta_\text{eff} /(2T_\text{eff})$
of the dissipative driving is proportional to the propulsion speed $v$
of the particle and $\zeta_\text{eff}$ and $T_\text{eff}$ denote the
appropriate effective hot Brownian friction and temperature parameters
for the Janus particle.  The exponent in Eq.~(\ref{eq:FR}) thus has
the interpretation of a non-equilibrium (hot Brownian) entropy
production due to dissipation to a virtual bath at the effective
temperature $T_\text{eff}$. The exact symmetry in Eq.~(\ref{eq:FR}) is
found to hold far from thermal equilibrium, even though the solvent is
not in thermal equilibrium and the driving is not due to a
deterministic external force, as usually assumed in the derivation of
Eq.~(\ref{eq:FR}). Instead, the hot Brownian swimmer is surrounded by
a temperature gradient and the driving is due to its thermophoretic
(force-free) self-propulsion.

\section{Molecular Dynamics Simulations}\label{sec:hotsim}
Numerical simulations of hot microswimmers can be performed on various
levels of coarse graining. One has to decide whether one can content
oneself with a numerical modeling of the swimming engine on the
phenomenological level, where it is subsumed into an effective slip
boundary condition as in the above theoretical discussion, or whether
higher resolution is required. In the former case, one can use
efficient strategies to solve the hydrodynamic flow patterns around
swimmers and between swimmers and other immersed bodies
\cite{Valadares2010,Yang2011a,Yang2013a,Fedosov2015}. In the second
case, classical atomistic molecular dynamics simulations (e.g.\ with
Lennard--Jones particles) are more suitable. Here, we pursue the
second route, since we want to be able to resolve some microscopic
details, such as the interfacial thermal resistance
(``Kapitza resistance'') and the mechanism of phoresis on an atomic
scale, which cannot be captured by effective thermodynamic or
hydrodynamic theories.

We simulate a spherical nanoparticle made of Lennard--Jones atoms that
are tightly bound together by a FENE potential
$U(r)=-0.5\kappa_{\alpha \beta} R_0^2 \ln (1-(r/R_0)^2)$, with
$R_0=1.5 \sigma$, which is immersed in a Lennard--Jones solvent.  A
simple strategy to atomistically realize the double-faced structure of
the experimentally employed Janus particles is to give the atoms on
the two hemispheres different thermal resistances to the solvent
\cite{Schachoff-etal:2015}. (This is easier to achieve than maintaining
a strong temperature gradient inside the particle by asymmetric
heating or making the thermophoretic mobility of its hemispheres
strongly heterogeneous. It is only the temperature gradient in the
solvent that matters, after all.) The Kapitza resistances, in turn, are
very sensitive to the wetting properties of the particle surface that
can, in a simple way, be encoded in the atomic particle-solvent
interaction potentials, given by the modified Lennard--Jones $12-6$
potential,
\begin{equation}
  U_{\alpha \beta}(r)=4 \epsilon \left[\left(\sigma/r \right)^{12}
  -c_{\alpha \beta}\left(\sigma/r\right)^6\right]\,,
\end{equation}
with an interaction cutoff at $r=2.5 \sigma$.  Here $c_{\alpha
  \beta}$ play the role of wetting parameters for the various atom
types $\alpha$ and $\beta$ \cite{Barrat1999,Barrat:2003}. The value
$c_\mathrm{ss}=1$ corresponds to the standard Lennard--Jones
interaction, which we choose for the mutual interactions between the
solvent particles. The atoms in a boundary layer of thickness $\approx
1\sigma$ on one hemisphere of the nanoparticle represent the gold cap
and are characterized by $c_\mathrm{gs}$ while the bulk atoms
represent the polystyrene core of the particle and are characterized
by $c_\mathrm{ps}$.  For the FENE spring constants, we use
$\kappa_\mathrm{gg}=30 \epsilon/\sigma^2$ and
$\kappa_\mathrm{gp}=\kappa_\mathrm{pp}=35\epsilon/\sigma^2$.  Choosing
different wetting parameters has the effect of varying the minimum
position $(2\sigma^6/c_{\alpha \beta})^{1/6}$ of the pair
potential. For $c_{\alpha \beta}=2$, the equilibrium distance between
the centers of a particle in the colloid and the solvent is $\sigma$,
for $c_{\alpha \beta}\to 0$ the attractive part and the local minimum
of the potential are absent.

A typical simulation run consists of an equilibration phase in the NPT
ensemble, with a Nos\'e--Hoover thermostat and barostat, at a
temperature of $T_0=0.75 \epsilon/k_{\rm B}$ and a thermodynamic
pressure of $p=0.01 \sigma^3/\epsilon$.  In the ensuing heating phase,
the global thermostat is then switched off and a non-equilibrium
steady state is created: (a) by thermostating a vertical domain of
solvent particles at the center of the simulation box to establish a
tent-shaped temperature field (phoresis), or (b) by thermostating the
``gold cap'' at the temperature $T_\mathrm{p}$ by a momentum
conserving velocity rescaling procedure (self-phoresis).  The fluid at
the boundary of the simulation box is always kept at the ambient
temperature $T_0$, by a similar rescaling procedure. Data acquisition
starts once the system has reached a steady state.

\begin{figure}[t]
\begin{tikzpicture}
\node at (-50pt, 65pt) {{\it (a)}};
\node at (75pt, 65pt) {{ \it (b)}};
\node at (75pt, -20pt) {{\it (c)}};
\node at (155pt, 65pt) {{\it (d)}};
\node[above,right] (img) at (-2,0)
{\includegraphics[width=0.5\linewidth]{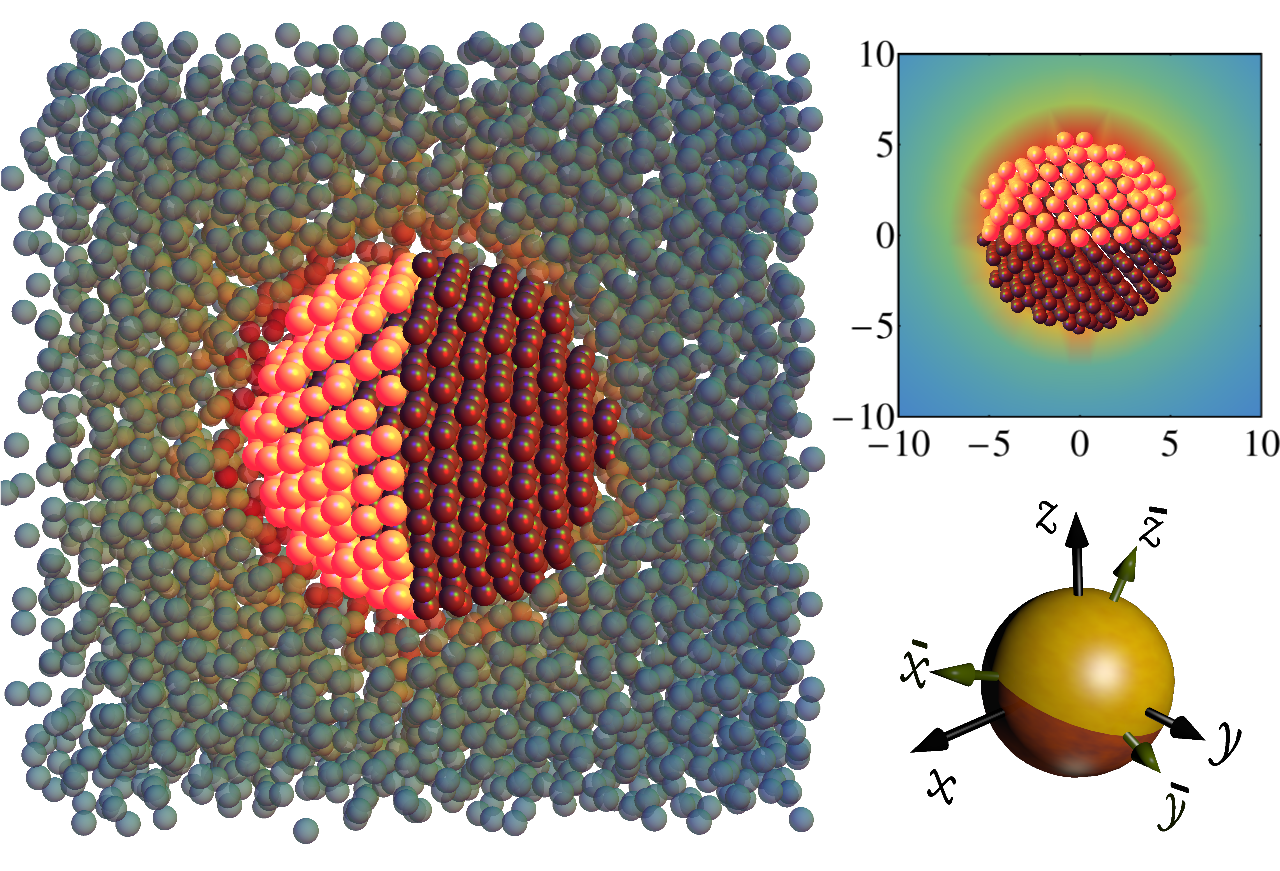}};
\node[above,right] (img) at (5,0)
{\includegraphics[width=0.4\linewidth]{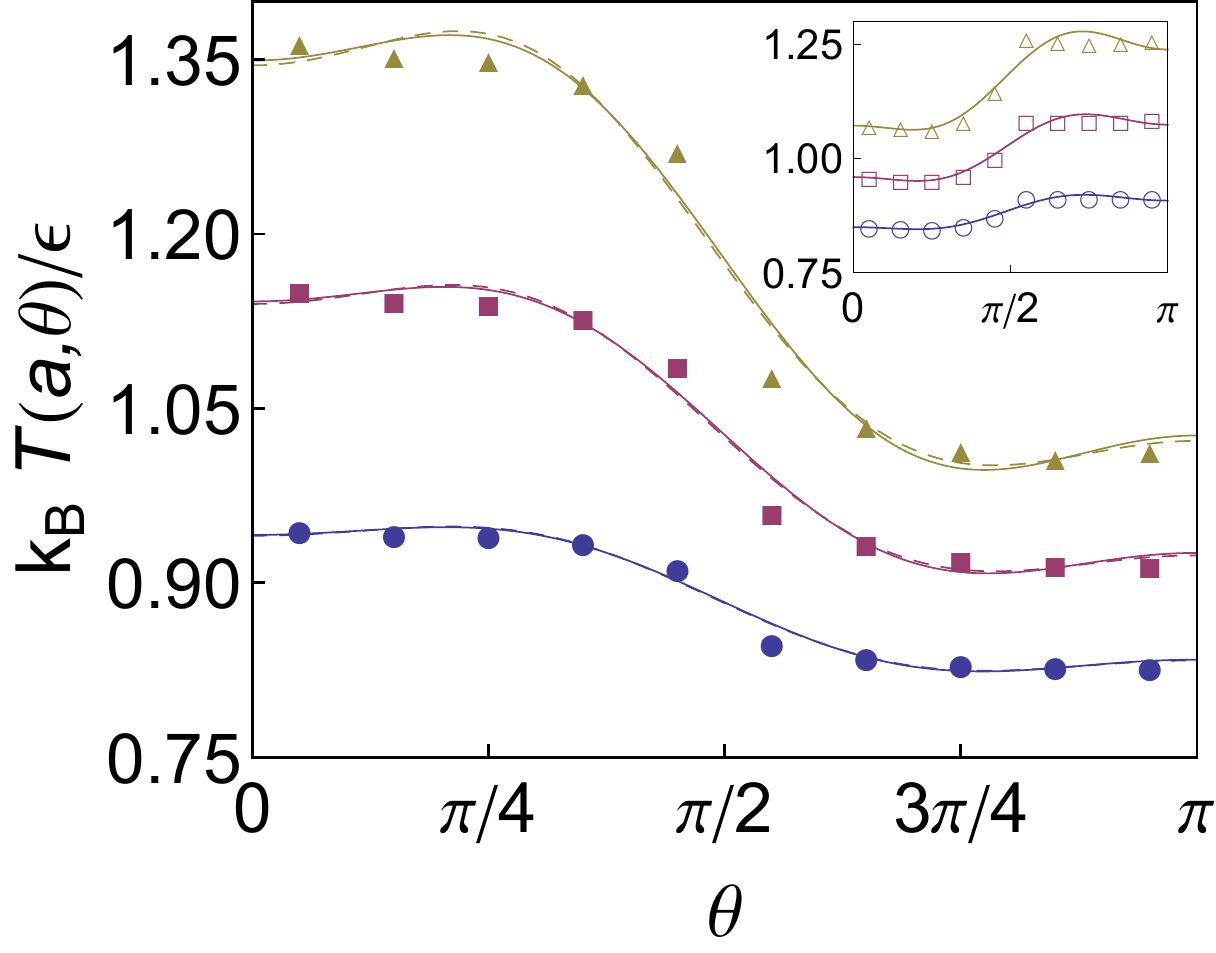}};
\end{tikzpicture}
\caption{Molecular Dynamics Simulations (adapted from
  Ref.~\cite{Schachoff-etal:2015}).  (a) Snapshot of a heated Janus
  particle with wetting parameters $c_\mathrm{gs}=2,~c_\mathrm{ps}=1$
  on the two hemispheres; coloring indicates the measured kinetic
  energy from which the continuum temperature field $T(r,\theta)$
  depicted in (b) is deduced. (c) Lab frame $(x,y,z)$ and
  (co-rotating) body frame $(\bar{x},\bar{y},\bar{z})$. (d) Solvent
  temperature $T(a,\theta)$ in a thin shell of thickness $0.5\sigma$
  around the heated Janus particle, with its cap maintained
  at temperatures $T_\mathrm{p}=1.20 \epsilon/\kb $ ({\Large
    \color{myblue} $\bullet$},{\Large \color{myblue} $\circ$}); $1.50
  \epsilon/\kb$ ({\small \color{mypurple} $\blacksquare$},{\small
    \color{mypurple} $\square$}) and $2.00 \epsilon/\kb$
  ({\color{myokker} $\blacktriangle$},{\scriptsize \color{myokker}
    $\triangle$}), for wetting parameters $c_\mathrm{gs}=2$,
  $c_\mathrm{ps}=1$ and $c_\mathrm{gs}=1$, $c_\mathrm{ps}=2$
  (inset). Solid lines represent fits by the series expansion from
  Eq.~(\ref{eq:heat_eq_soln}) and dashed lines solutions of the heat
  equation with the appropriate temperature-dependent thermal
  conductivity \cite{Chakraborty:2011}, both truncated after $n=3$
  (which causes spurious oscillations). }
  \label{fig:ch5_Janus_bead}
\end{figure}
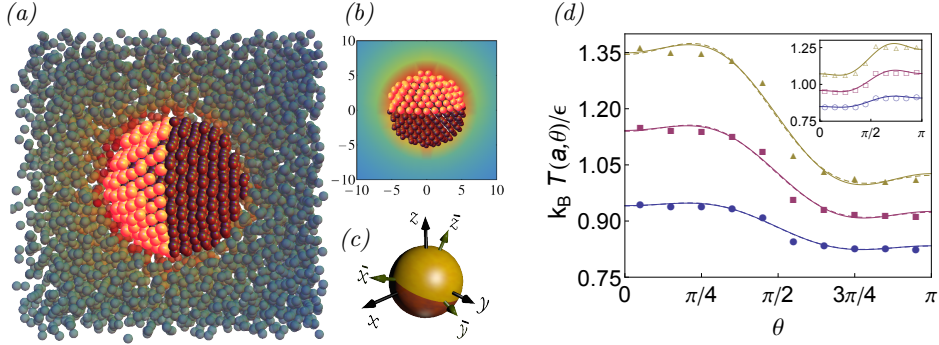

The temperature profile at the surface of the Janus particle can
directly be inferred from the average kinetic energy of nearby solvent
particles.  We parameterize the ensuing angular temperature variation
in the particle frame by a series of Legendre polynomials $P_n$,
\begin{equation}
  \label{eq:heat_eq_soln}
  T(a,\theta)=\bar T \sum_n B_n P_n(\cos \theta) \,,
\end{equation}
where $\bar T$ is the average temperature of the shell.  Truncating
the series at $n=3$ gives rise to some spurious oscillations, but
provides a decent description of the data
(Fig.~\ref{fig:ch5_Janus_bead}) and is also close to the theoretical
prediction from Fourier's heat equation for a previously established
temperature-dependence of the thermal conductivity of the
Lennard--Jones solvent \cite{Chakraborty:2011} (truncated to the same order).

\begin{figure}
\begin{tikzpicture}
\node at (-40pt, 70pt) {{\it (a)}};
\node at (145pt, 70pt) {{ \it (b)}};
%\node at (-40pt, -90pt) {{\it (c)}};
%\node at (155pt, -90pt) {{\it (d)}};
\node[above,right] (img) at (-2,0)
{\includegraphics[width=0.45\linewidth]{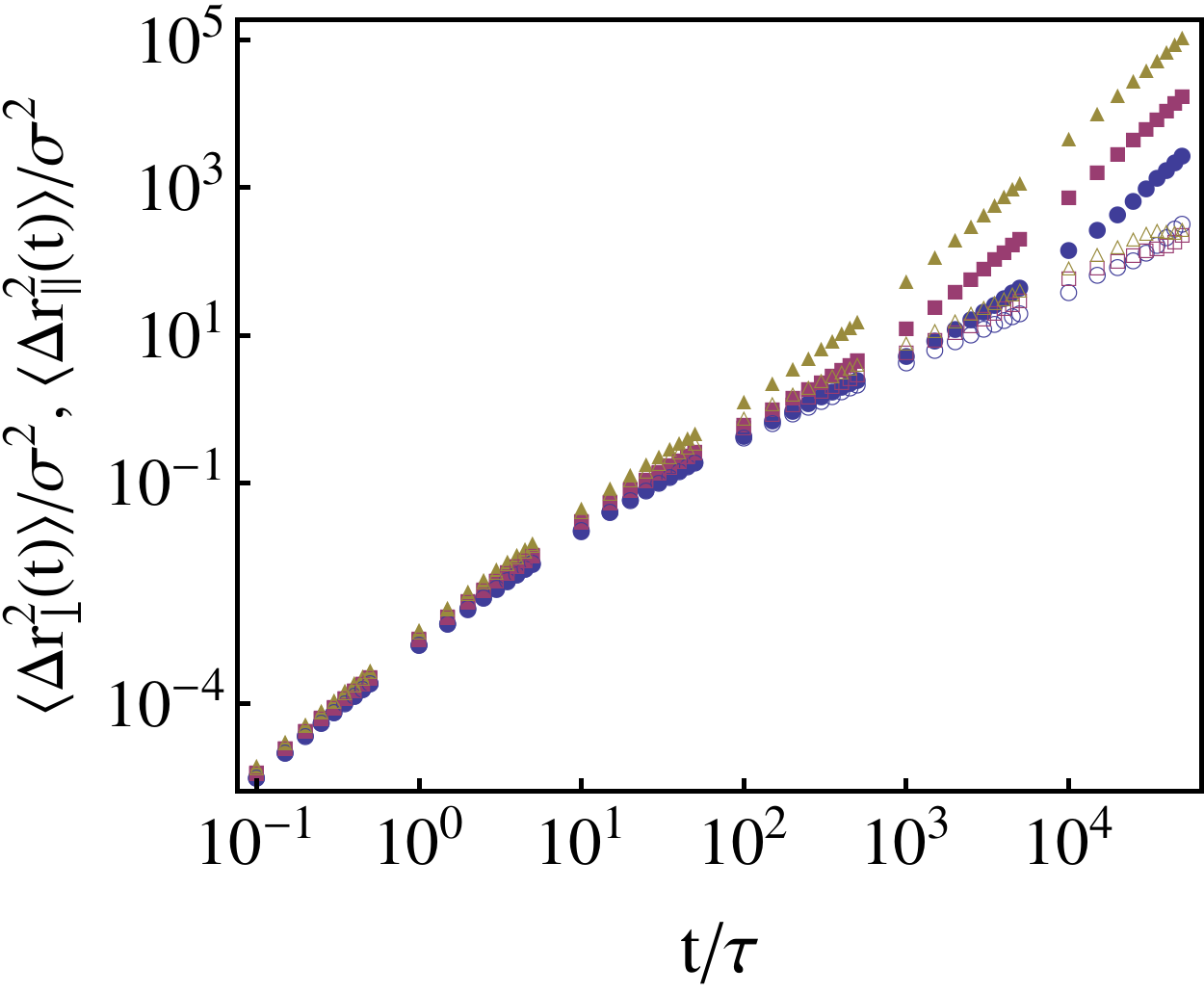}};
% \node[above,right] (img) at (5,0)
% {\includegraphics[width=0.5\linewidth]{images/soret_coeff_all_comp_linear_axis_cp1_cg2}};
% \node[below,left] (img) at (4.7,-6)
% {\includegraphics[width=0.5\linewidth]{images/soret_coeff_janus_comp_linear_axis}};
\node[below,right] (img) at (4.5,0)
{\includegraphics[width=0.45\linewidth]{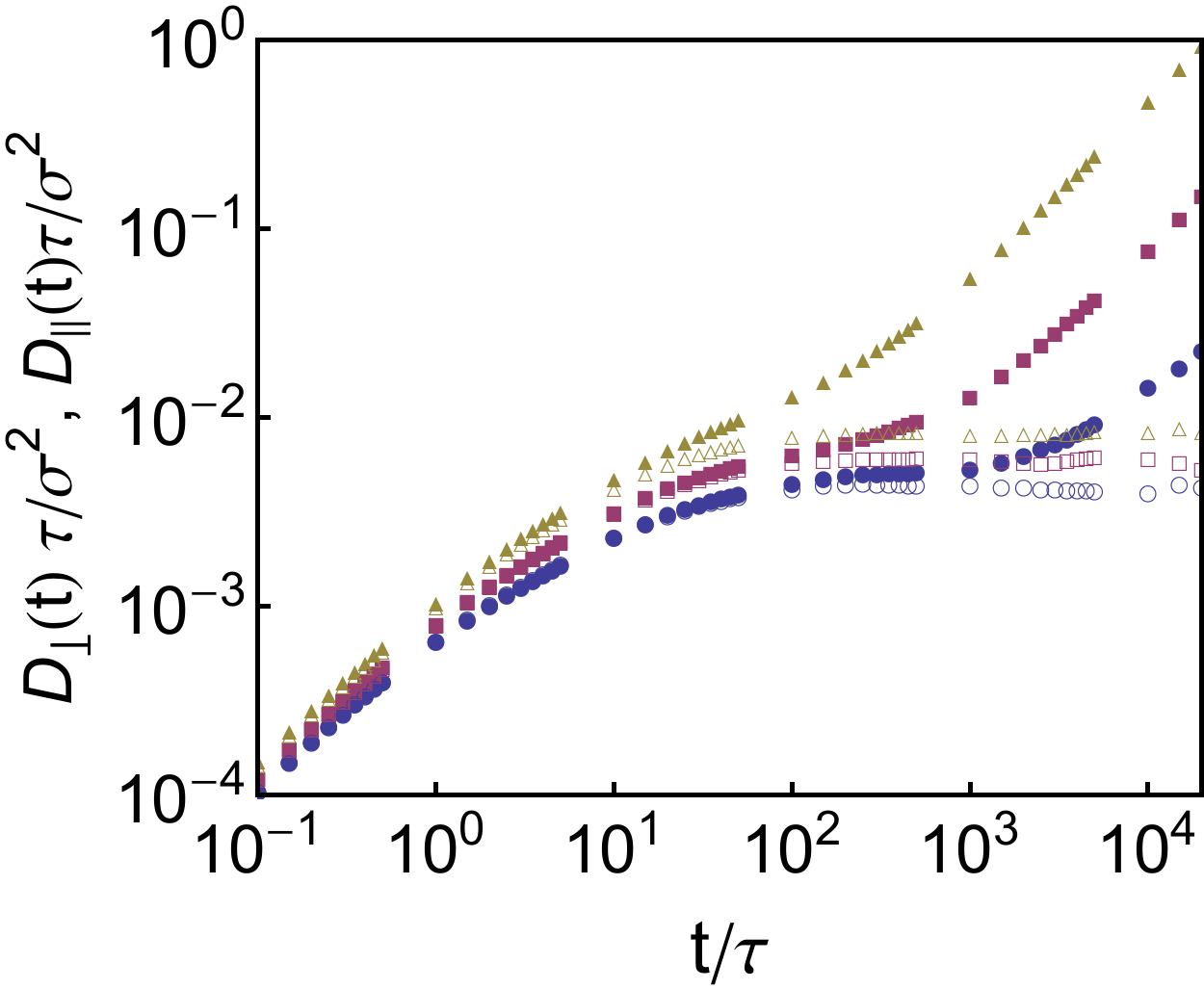}};
% \node[below,right] (img) at (7.5,0)
% {\includegraphics[width=0.35\linewidth]{images/janusdiffcomp}};
\end{tikzpicture}
 \caption{Mean-square displacements (a) and time-dependent
   diffusivities (b) along (filled symbols) and perpendicular to (empty
   symbols) the propulsion direction of a heated Janus particle
   ($c_\mathrm{gs}=2$ and $c_\mathrm{ps}=1$) in the particle frame.
   The temperatures of the hot cap are
   $T_p=1.10~\epsilon/\kb$~({\color{myblue} {\Large $\circ$,
       $\bullet$}}), $1.50~\epsilon/\kb$~({\color{mypurple} {\small
       $\square$,$\blacksquare$}}) and
   $2.00~\epsilon/\kb$~({\color{myokker}
     {$\vartriangle$,$\blacktriangle$}}). Adapted from
   Ref.~\cite{Schachoff-etal:2015}.}
  \label{fig:msd_janus}
\end{figure}

To obtain the data shown in the following figures, the velocity of the
Janus particle was measured both in the lab frame and in the co-moving
body frame.  The displacements and the velocities in the body frame
were obtained by projecting the corresponding quantities from the lab
frame at every time step of the simulations. At late times, the
mean-square displacements along and perpendicular to the instantaneous
propulsion direction of the self-thermophoretic Janus particle deviate
from each other, as expected from the superposition of diffusion and
ballistic self propulsion (Fig.~\ref{fig:msd_janus}). In the body
frame, the propulsion velocity can thus be read off from the
asymptotic slope of the mean-square displacement along the propulsion
direction (or from the corresponding velocity distribution).  In the
lab frame, the randomization of the propulsion direction due to the
rotational diffusion on the characteristic rotational diffusion time
scale $\tau_R$ ultimately renders the particle motion diffusive. (A
quantitative prediction for $\tau_R$ follows from the theory of hot
Brownian motion, as discussed above.)  The mean-squared displacement
as function of time (Fig.~\ref{fig:msd_janus})
\begin{equation}\label{eq:msd}
	\langle\Delta \vec r(t)^{2}\rangle=4 D
        t+2v^{2}\tau_{R}^{2}\left [ 
            t/\tau_{R}+e^{- t /\tau_{R}}-1\right]
\end{equation}
thus exhibits a characteristic crossover from ballistic motion at
short times $t\ll\tau_R$ to diffusive motion with an effective
diffusion coefficient of $D_{\rm eff}=D+v^{2}\tau_{R}/2$ at late times
$t\gg \tau_{R}$.

\begin{figure}
\begin{tikzpicture}
\node at (-40pt, 80pt) {{\it (a)}};
\node at (138pt, 80pt) {{ \it (b)}};
\node[above,right] (img) at (-2,0)
%{\includegraphics[width=0.46\linewidth]{images/soret_coeff_all_comp_linear_axis_cp2_cg1}};
{\includegraphics[width=0.46\linewidth]{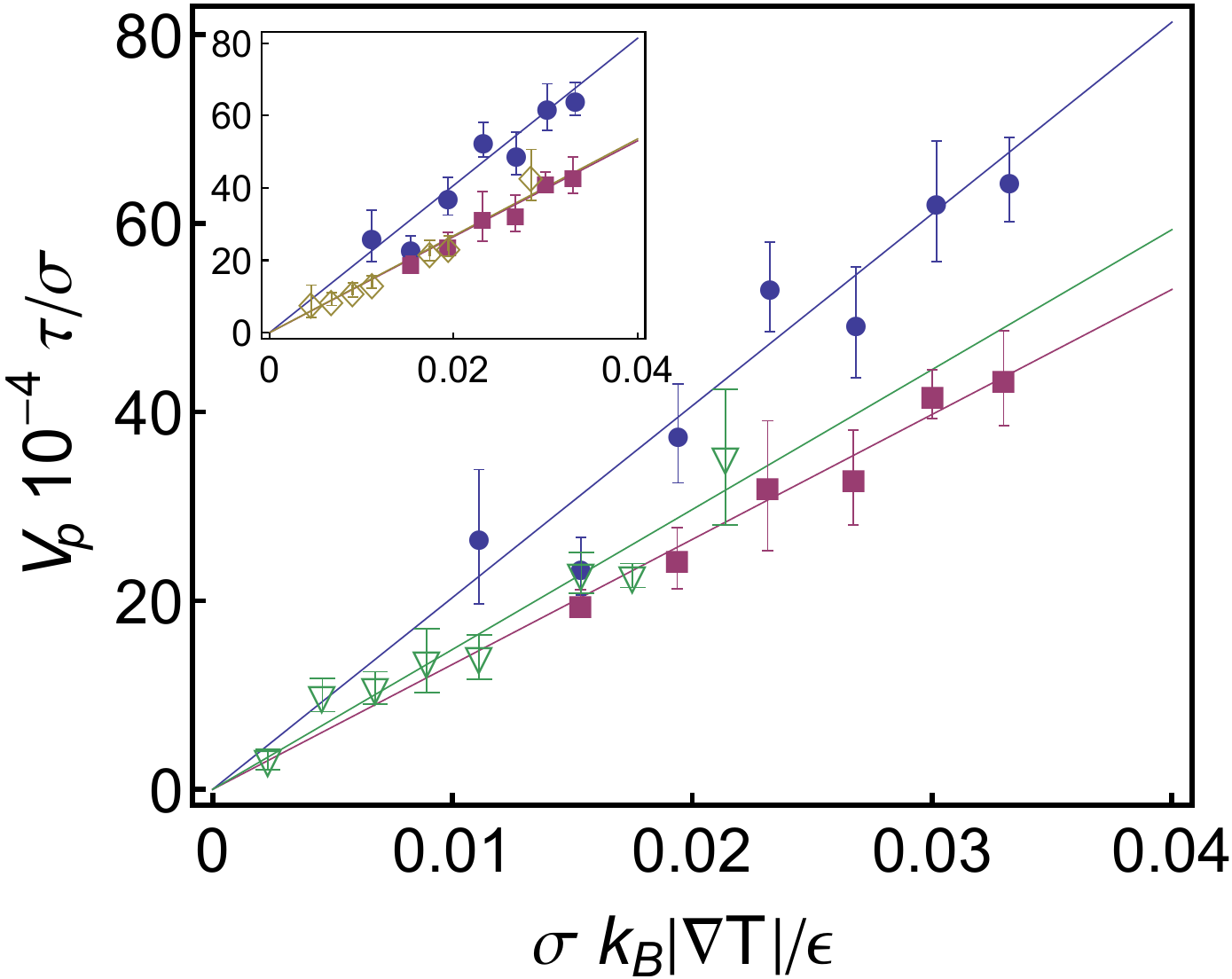}};
\node[below,right] (img) at (4.1,-0.08)
{\includegraphics[width=0.51\linewidth]{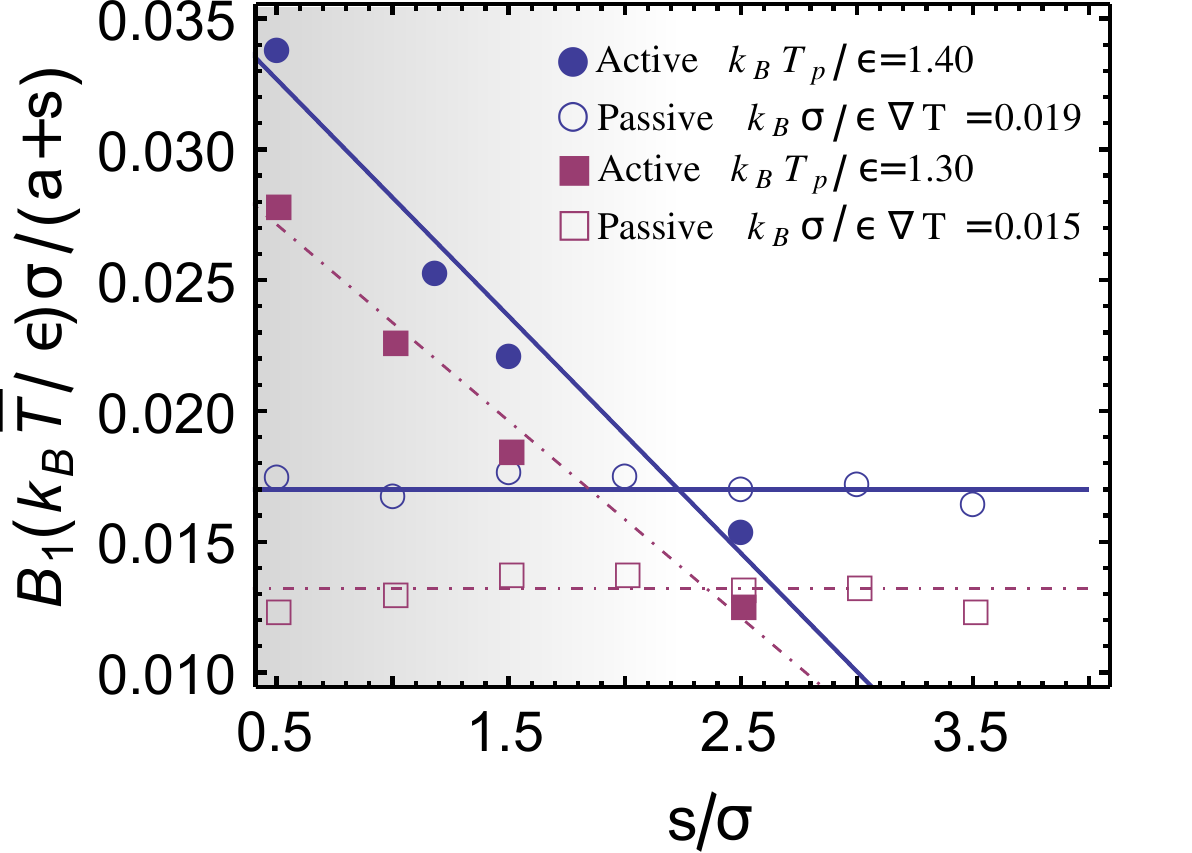}};
\end{tikzpicture}
\caption{Passive and active propulsion velocities for homogeneous
  beads and Janus beads. (a) Passive phoretic velocities for a
  homogeneous particle with $c_\mathrm{ps}=c_\mathrm{gs}=1$ ({\Large
    \color{myblue} $\bullet$}) and $c_\mathrm{gs}=c_\mathrm{ps}=2$
  ({\small \color{mypurple} $\blacksquare$}) and a Janus particle with
  $c_\mathrm{ps}=2, c_\mathrm{gs}=1$ ({\color{mygreen}
    $\triangledown$}) and $c_\mathrm{ps}=1, c_\mathrm{gs}=2$ ({\large
    \color{myokker} $\diamond$}) (inset).  (b) The component $\bar T
  B_1/(a+s)$ of the surface temperature gradient that causes the
  propulsion, as measured in fluid shells of various thicknesses $s$
  around the Janus bead. The active and passive propulsion velocities
  are the same for each heating power/temperature gradient, therefore
  the measured $\bar T B_1/(a+s)$ should also coincide for
  Eq.~(\ref{eq:mueff}) to hold. Lines guide the eye in inferring the
  corresponding effective boundary layer thickness $s$ (gray bar). }
  \label{fig:velocity_comparison}
\end{figure}

It is interesting to compare the propulsion velocities of Janus
particles moving in their self-generated temperature gradient to those
in an external temperature gradient.  In the passive phoretic
setup, the Janus particle was exposed to a constant temperature
gradient along the z-direction of the lab frame. In order to avoid a
temperature discontinuity across the periodic simulation boundaries,
the temperature profile was actually chosen to be tent-shaped around a
central maximum, and particle velocities to the left and right were
recorded and correlated separately. Moreover, to quantitatively
compare active and passive phoresis, the symmetry axis of the Janus
particle was subjected to an angular confinement during passive
thermophoresis, so that it remained parallel to the direction of the
external gradient. Altogether, four cases were investigated: a
homogeneous particle with homogeneous wetting parameters
$c_\mathrm{gs}=c_\mathrm{ps}=1$ and $c_\mathrm{gs}=c_\mathrm{ps}=2$
and a Janus particle with $c_\mathrm{gs}=1,~c_\mathrm{ps}=2$ and
$c_\mathrm{gs}=2,~c_\mathrm{ps}=1$, respectively.  In all four
scenarios we observed that the particle moved towards the cold,
indicating a positive phoretic mobility
(Fig.~\ref{fig:velocity_comparison}a).  Stronger potential attractions
correlate with weaker thermophoresis for the homogeneous
particles. The measured passive phoretic mobilities of the Janus
particle are intermediate between those of the hemispheres and
corroborate this trend.

Within the boundary layer approximation, the propulsion velocities for
both the passive and the active scenarios is determined by the
temperature on the surface of the particle and can be calculated from
Eqs.~(\ref{eq:slip_velocity}), (\ref{eq:propulsion_v}). In the
simulation, we can discern the finite thickness $s$ of the boundary layer,
and find that the temperature changes radially within it. This
prohibits a literal application of the boundary layer
equations. Instead, we propose here to subsume the two different
phoretic mobilities characterizing the two hemispheres of a Janus bead
into an effective mobility $\mu_{\rm eff}$, which is taken to be a
material parameter of the particle-solvent interface as a whole. The
propulsion speeds $v$ in both the active and the passive scenario are
then related to the relevant part of the temperature
gradient  $\bar T B_1/(a+s)$ (which excites the sliplet) by
\begin{equation}\label{eq:mueff}
v=- |\vec v_\text{tp}|=-\frac{2}{3} \mu_{\rm eff} \bar T B_1/(a+s) \;.
\end{equation}
By averaging the temperature field over shells of various thicknesses
for an active and a passive Janus particle moving at the same
propulsion velocity in the simulation, the effective boundary layer
thickness $s$ can thus be estimated by requiring this equation to hold
for both of them, simultaneously. From the data in
Fig.~\ref{fig:velocity_comparison}(b), $s$ is thereby found to be on
the order of the Lennard--Jones interaction range.

\section{``Hot'' Experimental Techniques}\label{sec:hotexp}
\begin{figure}
    \centering
     \includegraphics[width=0.9\hsize]{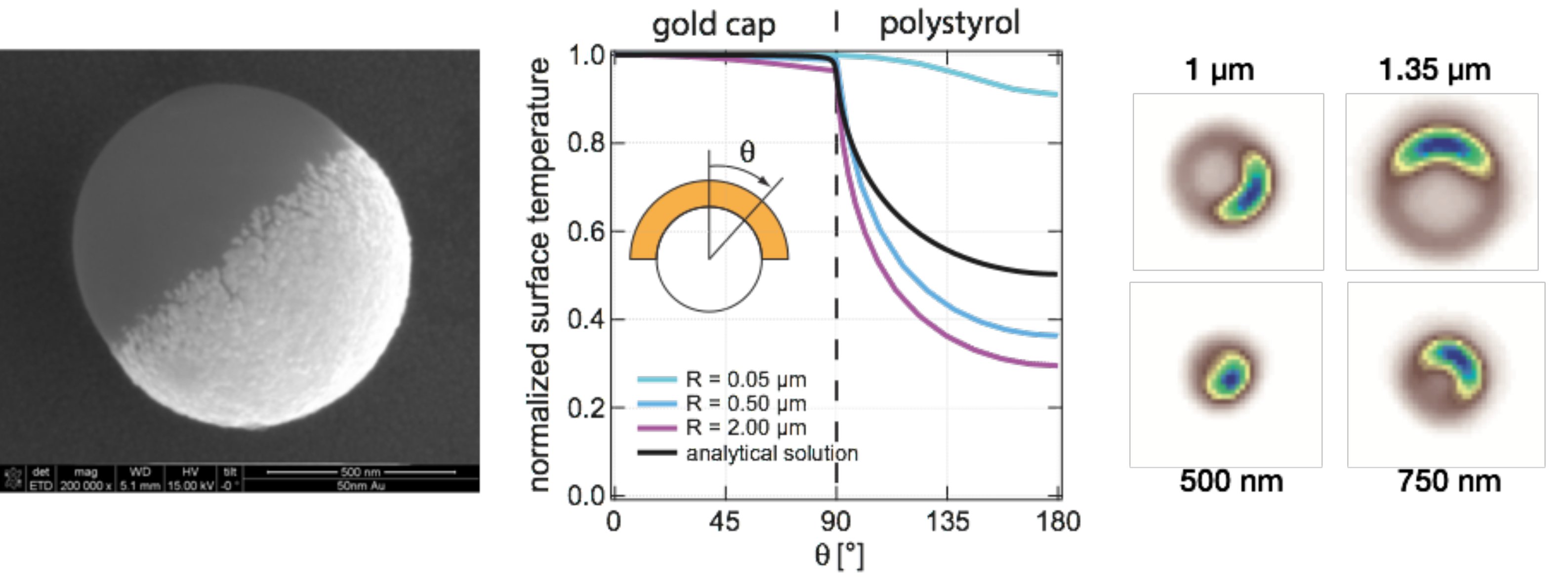}
     \caption{Electron microscopy image of a $1\mu m$ polystyrene
       particle covered with a 50 nm gold film on one hemisphere
       (left); finite-element simulation of the temperature field
       along a meridian and its analytical estimate
       \cite{bickel-majee-wuerger:2013} for an idealized model with an
       isothermal cap and assuming equal thermal conductivities for
       bulk and solvent (middle); false-color dark-field microscopy
       images of Janus particles of different radius (right), blue
       regions indicating the strongly scattering gold hemisphere. The
       in-plane orientation is determined by the image asymmetry. The
       out-of-plane orientation is obtained from the total intensity.
       Figure adapted from Ref.
       \cite{cichos2015fd2015}. }\label{fig:figure2}
\end{figure}

\textbf{Mechanism and synthesis of hot swimmers:} Hot swimmers can
very conveniently be fueled by absorbed light
\cite{jiang-yoshinaga-sano:2010} or by an oscillating magnetic field
\cite{baraban:2013}. In principle, also chemical energy could be used
to generate heat, but the ensuing diffusiophoretic effects would
likely largely overshadow the temperature effects
\cite{howse2007,Popescu_EPJST}. Note that the heating mechanism itself does not
provide a gradient, as in phoresis. It is the asymmetric design of the
particle that determines its propulsion direction.  In practice,
self-thermophoretic swimmers can for example be fabricated by partly
covering a small plastic or glass bead by a more strongly absorbing
material e.g.\ gold, or a number of other synthesis procedures, and a
variety of more complicated (e.g.\ chiral) designs are feasible
\cite{Fischer_EPJST}, which give rise to more fancy swimming
styles \cite{Loewen_EPJST}. As a simple and common example,
Fig.~\ref{fig:figure2} shows an electron microscopy image of a
gold-coated polystyrene particle. The gold cap acts as the asymmetric
heat source upon homogeneous illumination of the particle.  The
resulting temperature profile is depicted in Fig.~\ref{fig:figure2}
(central panel), as calculated by finite-element simulations for a
polystyrene particle with a 50 nm gold cap in water.

\begin{figure}
    \centering
   \includegraphics[width=0.9\hsize]{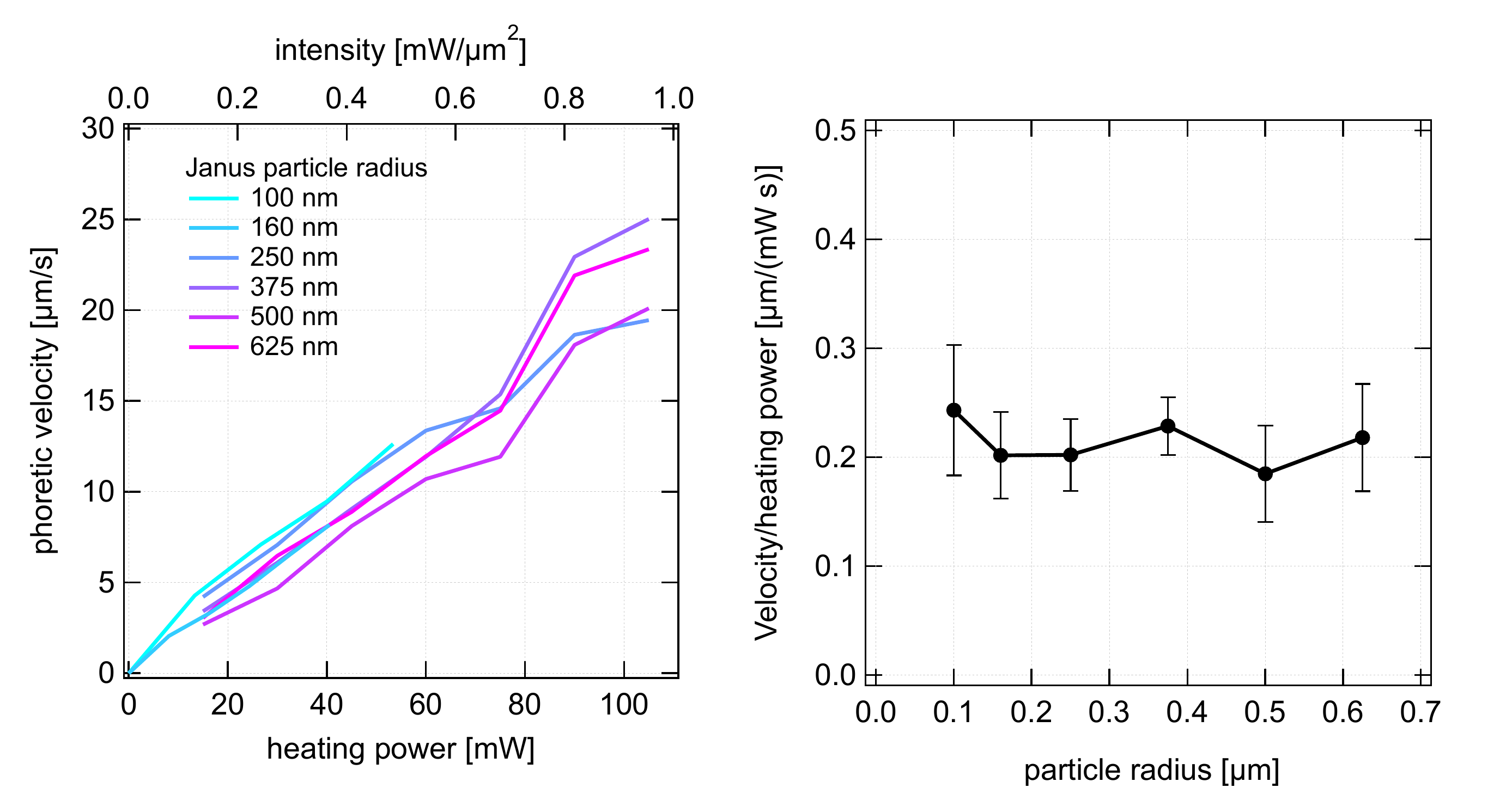}
     \caption{Dependence of the propulsion velocity of Janus particles
       on the absorbed heating power $P_{\rm abs}$ (left) and the particle radius
        (right). The particles all have a gold cap of 50 nm
       thickness but vary in size. Figure adapted from Ref. \cite{cichos2015fd2015}}\label{fig:figure3}
\end{figure}

\textbf{Propulsion velocity:} In a typical experiment, the gold-coated
hemisphere of a Janus particle is heated by the absorption of a
532\,nm laser.  Self-thermophoresis by laser heating exhibits two
characteristic properties illustrated by the experimental results in
Figure \ref{fig:figure3}: firstly, a linear dependence of the velocity
on the incident laser power (i.e.\ heating power); and secondly a
propulsion velocity that is independent of the particle size. Both can
be understood from simple arguments. As discussed in the context of
Eq.~(\ref{eq:sliplet}) and Eq.~(\ref{eq:propulsion_v}), the
thermophoretic propulsion velocity $\vec{v}_\text{tp}$ is the surface
average of the negative tangential slip velocity at the particle
surface, which is in turn proportional to the temperature gradient
$\vec{\nabla}_{||} T$ across the particle and its thermal mobility
coefficient $\mu(\vec{r})$
\cite{jiang-yoshinaga-sano:2010,Bickel:PhysicalReviewE:2014}.  The
thermophoretic propulsion velocity is thus directly proportional to
the temperature gradient. The tangential temperature gradient
$\vec{\nabla}_{||} T$ is proportional to the temperature jump $\delta
T$ divided by the particle radius $a$, i.e., $\vec{\nabla}_{||} T
\simeq \delta T/a$. The temperature jump $\delta T$ itself is
proportional to the power $P_{\rm abs}=\sigma_{\rm abs} I_{\rm inc}$
absorbed in the gold cap, where $\sigma_{\rm abs}$ is the absorption
cross section of the gold cap and $I_{\rm inc}$ is the incident laser
intensity. Altogether, we thus have
\begin{equation}
  \vec{\nabla}_{||}T\simeq \delta T/a=\frac{\sigma_{\rm abs}I_{\rm inc}}{4\pi\kappa_{\rm T} a^2},
\label{eq:DT}
\end{equation}
with an effective heat conductivity $\kappa_{\rm T}$.  One thus
arrives at the conclusion that the propulsion velocity is directly
proportional to the incident laser intensity $I_{\rm inc}$ and
independent of the particle radius $a$, if the absorption cross
section scales with the square of the particle radius,
$\sigma_\text{abs} \propto a^2$.  This is indeed the case for typical
microswimmer designs, if the thickness of the gold cap is independent
of the particle size, because the absorption cross section of a
micron-sized thin gold cap scales with the volume of the cap. For very
small particle sizes, additional considerations are required
\cite{cichos2015fd2015}.

\textbf{Detection methods:} To experimentally study hot swimmers and
hot Brownian motion, optical microscopy techniques are very
suitable. The paths of individual microswimmers can conveniently be
analyzed with the help of optical microscopy (bright field or
dark-field microscopy) under constant optical heating of the absorbing
gold hemisphere
\cite{Qian:ChemicalScience:2013,bregulla-yang-cichos:2014}. The
tracking concerns two observables: the 3-dimensional orientation of
the Janus particle and its position in the lab frame in each
exposure. Both are best determined in darkfield microscopy
\cite{Qian:ChemicalScience:2013} as the light scattering from a metal
cap and a polymer particle differ strongly (see
Fig.~\ref{fig:figure2}).  From a series of images one can calculate
the displacement vector between images separated by multiples of the
inverse framerate \cite{bregulla-yang-cichos:2014}. The
mean-squared displacement during that time period yields the effective
diffusion coefficient of that particle, while the mean displacement
vector in the particle frame yields the propulsion velocity.

A class of optical microscopy methods that put the heating to good use
for detection, and can thereby detect even nanometer-sized particles
and single molecules, are so-called photothermal techniques.  They
build on photothermal single-particle detection methods
\cite{Boyer2002a,berciaud-etal:2004}.  In case of a gold-capped Janus
particle, the illumination excites the conduction-band electrons in
the gold, which transfer their excitation energy to phonons within
some 100 femtoseconds. The heat is then released to the surrounding
liquid. A steady-state temperature profile in the solvent is quickly
(within a few microseconds) established by heat diffusion, for which
the solvent acts as an infinite heat bath. The basic idea of
photothermal microscopy is then to exploit the ``mirage'', i.e., the
induced refractive index change, around the heat source.  It acts as a
lens that can be detected by another laser, called the probe laser,
which is focused into the volume illuminated by the heating laser
\cite{berciaud-etal:2004,selmke2012photothermal}.  If a particle
diffuses through the focal volume of such a photothermal microscopy
setup, the length of the signal bursts gives information on the drift
and diffusion of the particle.  The method can be turned into a
photothermal correlation spectroscopy (PhoCS) technique
\cite{Octeau:AcsNano:2009,Paulo:JPhysChemC:2009,Raduenz2009}, which is
largely equivalent to fluorescence correlation spectroscopy
\cite{Dertinger:ChemPhysChem:2007,Schwille:BiophysJ:1999}. The main
difference is that the optical contrast is not caused by the probe
fluorescence but by the emanating heat. The experimental results
displayed in the left panel of Fig. \ref{fig:figure_HBM} have been obtained by a
slightly more sophisticated, highly quantitative (twin-focus)
implementation of this technique \cite{selmke2013twin}.

\textbf{Steering\index{Steering} by photon nudging:\index{Photon
    Nudging}} As mentioned above, the orientation of the Janus
particles symmetry axis is also subject to Brownian fluctuations,
which randomize the direction of propulsion. Therefore, microswimmers,
and nanoswimmers only retain their ballistic trajectories for times
shorter than the rotational diffusion time. As this rotational
diffusion time scales with the particle radius cubed, small particles
loose their directionality already after a few 100 microseconds and
reveal an enhanced diffusive motion rather than a ballistic
propulsion. A certain degree of built-in persistence of the swimming
motion is crucial for applications. To this end, two strategies can be
pursued.  One can slow down rotational diffusion by designing slender
swimmer bodies \cite{Sanchez_EPJST}.  Or one can employ feedback-based
control mechanisms to rectify the orientational fluctuations and yield
more directed motion or even perfectly steered motion
\cite{Qian:ChemicalScience:2013,bregulla-yang-cichos:2014}.

\begin{figure}
    \centering
   \includegraphics[width=0.9\hsize]{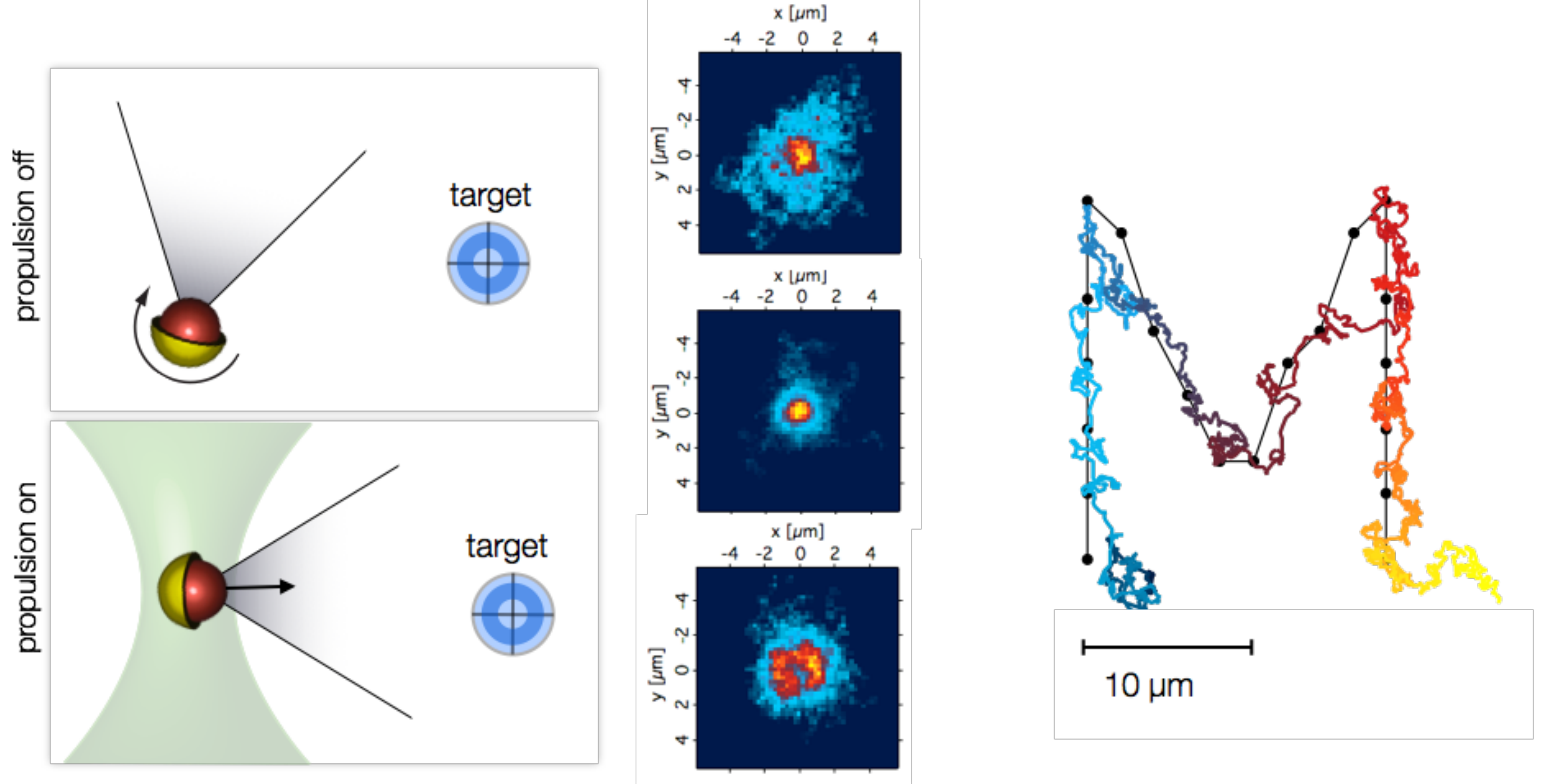}
   \caption{Photon nudging. \emph{Left:} sketch of the nudging
     principle; the propulsion is only switched on if the particle
     orientation is found within an acceptance angle of a desired
     target direction. \emph{Middle:} false-color histograms of
     particle positions around the target location; trapping is
     limited by diffusion for small nudging velocities and acceptance
     angles (top), diffusive and ballistic displacements are balanced
     for optimum localization (middle), and the particle overshoots
     the target for too rapid nudging or too large acceptance
     angles. \emph{Right:} nudging along a path defined by successive
     targets (black dots).}
\end{figure}

For the latter purpose, self-thermophoretic propulsion by laser
heating is well suited, as the propulsion can be switched on and off
at will. A successful technique that uses the switchable motion and
the information about the current orientation of the Janus swimmer to
steer and trap swimmers in solution is called ``photon nudging''
\cite{Qian:ChemicalScience:2013}. During phases of rotational Brownian
motion without self-propulsion, the orientation is analyzed in real
time. If the particle orientation is by chance found within a certain
acceptance angle $\theta$ around the desired target direction, the
propulsion is switched on.  This causes a net motion of the swimmer
towards the target location and finally a localization of the swimmer
at the target position.  Note that, for homogeneous illumination, this
type of steering or trapping does not involve any external forcing nor
any external torque (as rotational Brownian motion provides the random
reorientation mechanism). It is therefore reminiscent of a Maxwell
daemon.

The accuracy of localization by photon nudging is again limited by
diffusion, mostly during the off periods, when the particle goes off
track. Now, the passive rotational diffusion not only limits the
persistence of the path, and thus leads to intermittent rest phases of
the swimmer, it also achieves the reorientation during these phases
that initiates the next active phase. The naively expected growth of
the relative localization error (the square root of the mean-square
displacement of the particle from the target location divided by the
particle radius) with decreasing particle size can therefore be
avoided with the nudging method
\cite{bregulla-yang-cichos:2014,Qian:ChemicalScience:2013}.

\section{Conclusion}
We have introduced some basic notions and techniques relevant for
self-thermophoretic microswimmers and gathered some important recent
results concerning their hot Brownian fluctuations, their phoretic
boundary layers, and their experimental detection and active feedback
control. A more detailed investigation of the precise microscopic
conditions within the boundary layer would be a worthwhile task for
future numerical work.  Ongoing work moreover attempts to put these
techniques to good use for the analysis of mutually interacting
swimmers and swimmers interacting with liquid-solid boundaries
\cite{Stark_EPJST}. An important goal would be to establish
coarse-grained hydrodynamic or thermodynamic theories or even a
statistical mechanics of many-body swimming \cite{Speck_EPJST}, along
these lines.  Our discussion has indicated that one may expect several
new and interesting aspects to show up in such theories that are not
present in conventional many-body theories for passive particles,
chiefly due to the complicated non-equilibrium (and not pairwise
additive) interactions and to the non-equilibrium character of the
Brownian fluctuations of hot swimmers.  These unconventional effects,
if well understood and controlled, could possibly again inspire new
applications and techniques of manipulation, along the directions
outlined here.

\subsection{Acknowledgement}
\label{sec:2}
We acknowledge financial support for K.K. and F.C.  from the German
Science Foundation DFG, SPP1726, and for D.C.  from the Science and
Engineering Research Board, India vide grant no SB/S2/CMP-113/2013 and
from the Alexander von Humboldt Foundation for his stay in Leipzig in
2015.

\bibliographystyle{apsrev4-1}
\bibliography{extracted}

%\begin{thebibliography}{}
%Author, Journal \textbf{Volume}, (year) page numbers
%Author, \textit{Book title} (Publisher, place year) page numbers
% etc
%\end{thebibliography}

\end{document}